\documentclass[review]{elsarticle}

\usepackage{natbib}
\usepackage{lineno, hyperref}

\usepackage{wrapfig}

\usepackage{amssymb,amsmath}
\usepackage{bm,upgreek}
\usepackage{mathrsfs}
\usepackage{tablefootnote}
\usepackage{graphics}
\usepackage{amsfonts}
\usepackage{color}
\usepackage{ulem}
\graphicspath{{Fig3D/}{Fig/}}

\newcommand\rd{{\rm{d}}}
\newcommand{\bfm}[1]{{\rm\bf #1}}

\definecolor{myviolet}{RGB}{255,0,255}
\definecolor{darkgreen}{rgb}{0,0.5,0}

\usepackage{lipsum}
\makeatletter
\def\ps@pprintTitle{%
 \let\@oddhead\@empty
 \let\@evenhead\@empty
 \def\@oddfoot{\url{https://doi.org/10.1016/j.cma.2021.113705}}%
 \let\@evenfoot\@oddfoot}
\makeatother

\begin{document}

\begin{frontmatter}

\title{Phase-field modeling of multivariant martensitic transformation at finite-strain: computational aspects and large-scale finite-element simulations}

\author[FMP]{K. T\r{u}ma}
\ead{ktuma@karlin.mff.cuni.cz}

\author[IPPT]{M. Rezaee-Hajidehi}
\ead{mrezaee@ippt.pan.pl}

\author[FMP]{J. Hron}
\ead{hron@karlin.mff.cuni.cz}

\author[OXF]{P. E. Farrell}
\ead{patrick.farrell@maths.ox.ac.uk}

\author[IPPT]{S. Stupkiewicz\corref{cor1}}
\ead{sstupkie@ippt.pan.pl}

\cortext[cor1]{Corresponding author. Tel.: (+48) 22 826 12 81 ext.\ 338.}

\address[FMP]{Faculty of Mathematics and Physics, Charles University, Sokolovsk\'{a} 83, 186 75, Prague, Czech Republic}

\address[IPPT]{Institute of Fundamental Technological Research (IPPT), Polish Academy of Sciences,\\
Pawi\'nskiego 5B, 02-106 Warsaw, Poland}


\address[OXF]{Mathematical Institute, University of Oxford, Oxford OX2 6GG, UK}

\begin{abstract}
Large-scale 3D martensitic microstructure evolution problems are studied using a finite-element discretization of a finite-strain phase-field model.
The model admits an arbitrary crystallography of transformation and arbitrary elastic anisotropy of the phases, and incorporates Hencky-type elasticity, a penalty-regularized double-obstacle potential, and viscous dissipation. The finite-element discretization of the model is performed in Firedrake and relies on the PETSc solver library. The large systems of linear equations arising are efficiently solved using GMRES and a geometric multigrid preconditioner with a carefully chosen relaxation. The modeling capabilities are illustrated through a 3D simulation of the microstructure evolution in a  pseudoelastic CuAlNi single crystal during nano-indentation, with all six orthorhombic martensite variants taken into account. Robustness and a good parallel scaling performance have been demonstrated, with the problem size reaching 150 million degrees of freedom.
\end{abstract}


\end{frontmatter}

\section{Introduction}
The phase-field method has proven to be a powerful computational tool for modeling microstructure evolution in various material systems. The essential feature of the phase-field method is that the interfaces are assumed to be diffuse and accordingly the tremendous computational burden of tracking sharp interfaces is avoided. Thanks to its computational advantages, the phase-field method has been extensively employed in different areas of materials science and physics, e.g.~\cite{chen2002phase,moelans2008introduction, steinbach2009phase,wang2010phase,provatas2011phase}.

Modeling the martensitic phase transformation (as in shape memory alloys) constitutes one of the classical applications of the phase-field method. Successful studies in this genre include the seminal works of Khachaturyan and co-workers \cite{wang1997three,artemev2000three,jin2001three}, Chen and co-workers \cite{wen2000phase,li2001phase}, as well as the subsequent developments that followed, e.g.~\cite{levitas2002threeA, ahluwalia2004landau,shu2008multivariant,levitas2009displacive, lei2010austenite,hildebrand2012phase,she2013finite, borukhovich2014large,tuuma2016size}. A wide class of these studies is limited to the use of spectral solvers, e.g.~\cite{wang1997three, artemev2000three,wen2000phase,jin2001three,li2001phase, ahluwalia2004landau,shu2008multivariant,lei2010austenite, zhong2014phase,borukhovich2014large, zhao2020finite}, see also \cite{chen1998applications}. Although the models relying on FFT-based spectral solvers have definite advantages due to their high computational efficiency, they are restricted to problems with a periodic unit cell, and thus cannot deal with problems with arbitrary geometry and boundary conditions. Moreover, most of the models in this category are formulated within the small-strain regime, with only a limited number of them incorporating a finite-strain theory, e.g.~\cite{borukhovich2014large,zhao2020finite}. 

On the other hand, models discretized within the finite-element framework are not subject to the above limitations. In particular, they can cope with finite-strain formulations in a straightforward manner \cite{levitas2009displacive,clayton2011phase, hildebrand2012phase,tuuma2016size,bartels2017efficient, basak2019finite}. From the computational point of view, however, it is very well known that the FFT-based models are potentially more efficient compared to the finite-element models and are amenable to problems with very fine resolution for which the use of the finite-element models would be restrictive \cite{eisenlohr2013spectral}, see also \cite{schneider2017fft,zeman2017finite}. With this in mind, the goal that we pursue in the present work is to develop a robust and efficient computational model using finite elements and multigrid solvers for the phase-field modeling of multivariant martensitic transformation in shape memory alloys. The model is required to offer good parallel scaling performance so that it can be employed for large-scale simulations.

Martensitic phase transformation is a first-order solid-solid displacive transformation that occurs between a higher-symmetry phase (austenite) and lower-symmetry phase (martensite), and is characterized by microstructure evolution, which, for instance, in shape memory alloys provides the basic mechanism for striking properties such as pseudoelasticity and the shape memory effect \cite{Bhattacharya2003}. Of the numerous phase-field models developed to study microstructure evolution during martensitic phase transformation, some are considered within the framework of the finite-element method. Stress- and temperature-induced displacive transformations have been addressed in a number of studies, e.g.~\cite{levitas2010surface, yeddu2012three,schmitt2013phase, cui2017phase,mamivand2013phase,she2013finite,CISSE2021106144}, including those in the finite-strain setting, e.g.~\cite{levitas2009displacive,clayton2011phase, hildebrand2012phase,tuuma2016size,bartels2017efficient, basak2019finite}.

In the context of the phase-field approach, a sufficiently fine mesh resolution must be adopted to represent the interfaces of complex microstructure patterns. However, this is achieved at the expense of high computational cost and high memory requirements, which may lead to computational challenges, in particular, in more involved 3D problems.
Accordingly, finite-element simulations of microstructure evolution are typically restricted to 2D and relatively simple 3D problems, while more involved 3D problems are not common, e.g.~\cite{yeddu2012three,dhote20153d}. 
In order to overcome these limitations and improve modeling capabilities, attempts have been made to develop novel numerical strategies, such as adaptive mesh refinement \cite{mahnken2013goal}, isogeometric analysis \cite{dhote20153d}, statistical learning and optimization algorithms \cite{wei2020improved} or a multiscale approach \cite{kochmann2016two}. 
Nevertheless, it is still an ongoing challenge to develop finite-element-based phase-field models that can be efficiently employed for large-scale 3D microstructure evolution problems.
In this work we address this computational challenge through the use of a multigrid method with a carefully chosen relaxation that honors the strong coupling among order parameters.

The computational model in the present work develops a physically relevant description of multivariant martensitic phase transformation within the phase-field framework and a robust finite-element implementation. The finite-strain phase-field model admits an arbitrary crystallography of transformation (the cubic-to-orthorhombic transformation with 6 martensite variants is considered as an application) and an arbitrary anisotropy of phases (consistent with the symmetry of the phases). The constitutive description is based on the elastic strain energy of Hencky-type, e.g.~\cite{rezaee2020phase}, a multiphase double-obstacle potential \cite{steinbach2009phase} and a viscous-type dissipation potential. A variational formulation of the model is developed such that the complete evolution problem is governed by an incremental energy minimization \cite{hildebrand2012phase,tuuma2016size}. The coupled nonlinear equations resulting from the finite-element discretization are then solved by means of Newton's method. A 2D version of the present phase-field model has been previously introduced and used to study the martensitic transformation in 2D nano-indentation problems \cite{rezaee2020phase}. 

As already stressed, the main focus of the present work is to develop a finite-element-based computational model suitable for large-scale 3D problems. Accordingly, the finite-element implementation is performed in Firedrake \cite{rathgeber2016firedrake}, an automated finite-element package. Firedrake is tightly integrated with the PETSc solver library \cite{balay2019petsc}, and thereby provides an excellent platform for implementing the proposed discretization and multigrid preconditioner. In order to ease the burden on the iterative solver and to avoid further computational complexities associated with the use of Lagrange multipliers, the classical penalty regularization method is applied to treat the inequality constraints on the phase-field order parameters, and to enforce the contact condition in the indentation problem.

A pivotal component of the present model is the Hencky-type elastic strain energy that allows for an arbitrary anisotropy of phases. A usual and simple alternative would be the St.~Venant--Kirchhoff elastic strain energy. However, our initial analyses revealed that, because of the high compressive strains resulting from the contact problem considered in this study, the Hencky-type model performs much better than the St.~Venant--Kirchhoff model, see a detailed study in \cite{RezaeeHajidehi2020Pade}. The finite-element implementation of the Hencky-type model requires the computation of the matrix logarithm (elastic Hencky strain) and its first and second derivatives, which is not a straightforward task. For this purpose, the Pad\'e approximation method has been employed \cite{RezaeeHajidehi2020Pade}. A high-accuracy approximation of the matrix logarithm is then provided in an explicit formula, and, as a result, its derivatives can be calculated directly by the automatic differentiation algorithm in Firedrake.

As an application, we study the microstructure evolution in a CuAlNi single crystal during nano-indentation, along with its related pseudoelastic behaviour. The simulation results reveal interesting nontrivial microstructure characteristics, including the formation of complex patterns such as twinning and saw-tooth morphology. As far as we are aware, results of such scope and detail have not been reported in the literature so far. We carry out a parametric study to investigate the effect of various numerical parameters, and to verify the efficiency, robustness and scalability of the computational model.

\section{Model description}
The phase-field model presented in this study is adopted from our previous work \cite{rezaee2020phase}, see also \cite{tuuma2016size} for an earlier version of the model with two hierarchical order parameters for modeling austenite--twinned martensite microstructures. The model employs the elastic strain energy as a quadratic function of Hencky (logarithmic) strain, the multiphase double-obstacle potential, cf.~\cite{steinbach2009phase}, and a viscous-type dissipation potential, see Section~\ref{Sec-EnergyDiss}. A variational formulation of the model is developed, where the complete evolution problem is formulated in the incremental energy minimization framework, see Section~\ref{Sec-energyMin}.

\subsection{Order parameters and kinematics}
The current model considers a parent phase (austenite) and $N$ product phases (variants of martensite). Each phase $i$ is characterized by an order parameter $\eta_i$. The order parameters are individually bounded between 0 and 1, and are jointly subject to a sum-to-unity constraint, i.e.\
\begin{equation}\label{Eq-order}
0 \leq \eta_i, \quad \text{for} \quad i=0,\dots,N \qquad \text{and} \quad \sum_{i=0}^N \eta_i=1,
\end{equation}
where the above conditions imply also the fulfillment of the inequality constraints $\eta_i \leq 1$ (we discuss the treatment of the constraints in the finite-element procedure in Section~\ref{Sec-Penalty}). The order parameters can be interpreted as the phase volume fractions, and they are used to directly interpolate various material properties within the diffuse interfaces, in contrast to other approaches that rely on specially-designed interpolation functions, see e.g.\ \cite{levitas2018phase}.

The phase-field model is developed within the finite-deformation setting. The total deformation gradient $\bfm{F}=\nabla \bm{\upvarphi}$, where $\bm{\upvarphi}$ denotes the mapping from the reference to the current configuration and $\nabla$ denotes the gradient relative to the reference configuration, is multiplicatively split as
\begin{equation}
\bfm{F}=\bfm{F}^\text{e}\bfm{F}^\text{t}, \qquad 
\end{equation}
where $\bfm{F}^\text{e}$ and $\bfm{F}^\text{t}$ represent, respectively, the elastic and transformation parts of $\bfm{F}$. The transformation part $\bfm{F}^\text{t}$ is described as a linear mixture of the transformation stretch (Bain strain) tensors of individual phases $\bfm{U}_i^\text{t}$, i.e.\
\begin{equation}\label{Eq-LinMix}
\bfm{F}^\text{t}(\bm{\upeta})=\sum_{i=0}^N \eta_i \bfm{U}_i^\text{t}, \qquad \bm{\upeta}=(\upeta_0, \upeta_1,\dots,\upeta_N).
\end{equation}
Assuming undeformed austenite as the reference configuration, $\bfm{F}^\text{t}= \bfm{U}_0^\text{t}=\bfm{I}$ corresponds to the pure austenitic state, where $\bfm{I}$ is the second-order identity tensor, while the transformation stretches $\bfm{U}_i^\text{t}$ result from the crystallography of the phase transformation.

The transformation deformation gradient $\bfm{F}^\text{t}$ can be alternatively formulated with the logarithmic mixing rule, namely $\bfm{F}^\text{t}=\exp \Big( \sum_{i=0}^N \eta_i \log \bfm{U}_i^\text{t}\Big)$, cf.~\cite{tuuma2016size}. The basic feature of the logarithmic mixing rule with respect to the linear one, Eq.~\eqref{Eq-LinMix}, is that for a fixed volume fraction of austenite $\eta_0$, the transformation between the martensite variants induces no volume change, i.e.\ $\det \bfm{F}^\text{t}$ remains constant, which is of course desirable from a physical standpoint, see the related discussion in \cite{tuuma2016size}.
However, it has been observed that the model based on the logarithmic mixing rule may result in spurious stresses within the diffuse martensite--martensite interfaces that are higher than those in the case of the linear mixing rule \cite{basak2017interfacial}. In addition, the computer implementation of the model based on the logarithmic mixing rule is more difficult, as it involves computation of the tensor (matrix) exponential.

\subsection{Free energy function and dissipation potential}\label{Sec-EnergyDiss}
The Helmholtz free energy density consists of three constituents, namely the chemical energy $F_\text{chem}$, the elastic strain energy $F_\text{el}$ and the interfacial energy $F_\text{int}$, viz.\
\begin{equation}
F(\bfm{F},\bm{\upeta},\nabla \bm{\upeta})=F_\text{chem}(\bm{\upeta})+F_\text{el}(\bfm{F},\bm{\upeta})+F_\text{int}(\bm{\upeta},\nabla \bm{\upeta}).
\end{equation}
The chemical energy $F_\text{chem}$ is defined as the weighted sum of the chemical energies of individual phases, $F_i^0$,
\begin{equation}
F_\text{chem}(\bm{\upeta})=\sum_{i=0}^N \eta_i F_i^0.
\end{equation}

The elastic strain energy $F_\text{el}$ is defined as a quadratic function of the elastic Hencky (logarithmic) strain $\bfm{H}^\text{e}$ in the following form,
\begin{equation}\label{Eq-elasticEn}
F_\text{el}(\bfm{F},\bm{\upeta})=\frac{1}{2}(\det \bfm{F}^\text{t})\bfm{H}^\text{e} \cdot \mathbb{L}(\bm{\upeta})\bfm{H}^\text{e}, \qquad \bfm{H}^\text{e}=\frac{1}{2}\log \bfm{C}^\text{e}, \qquad \mathbb{L}(\bm{\upeta})=\sum_{i=0}^N \eta_i \mathbb{L}_i,
\end{equation}
where $\bfm{C}^\text{e}=(\bfm{F}^\text{e})^\text{T}\bfm{F}^\text{e}$ represents the elastic right Cauchy-Green tensor, $\bfm{F}^\text{e}=\bfm{F}(\bfm{F}^\text{t})^{-1}$, and $\mathbb{L}$ represents the average fourth-order elastic stiffness tensor, which is defined by applying a Voigt-like averaging scheme to the elastic stiffness tensors of individual phases, $\mathbb{L}_i$.

As an alternative for the Hencky strain energy, Eq.~\eqref{Eq-elasticEn}, the St.\ Venant--Kirchhoff model is often employed, in which the elastic strain energy is formulated as a quadratic function of the elastic Green strain tensor $\bfm{E}^\text{e}=\frac{1}{2}(\bfm{C}^\text{e}-\bfm{I})$, as e.g.\ in \cite{maciejewski2005elastic,tuuma2016size,tuuma2016phase}. However, it is known that the St.\ Venant--Kirchhoff strain energy exhibits poor performance under compression, which is due to the lack of important properties, in particular, rank-one convexity \cite{raoult1986non,schroder2010poly}. In the indentation problems addressed in the present study, high compressive stresses develop beneath the indenter and, thus, the St.\ Venant--Kirchhoff strain energy may not be a suitable choice (as confirmed by our preliminary studies). On the other hand, the elastic Hencky strain energy, Eq.~\eqref{Eq-elasticEn}, has a larger domain of rank-one convexity and performs well for a wider range of strains \cite{neff2016exponentiatedIII}. For a detailed discussion on this matter, the reader is referred to \cite{RezaeeHajidehi2020Pade}. 
Note that the range of strains in which the elastic strain energy behaves well must be large enough to encompass also the strains encountered during Newton iterations, and this range is larger than that corresponding to the converged solution.

The finite-element implementation of the Hencky strain energy includes computation of the matrix logarithm together with its first and second derivatives, respectively, for the stress and the tangent operator. Following \cite{RezaeeHajidehi2020Pade}, Pad\'e approximants are employed in this study to facilitate the implementation, see Section~\ref{Sec-HenckyApprox}.

Finally, the interfacial energy $F_\text{int}$ is adopted in the double-obstacle form \cite{steinbach2009phase},
\begin{equation}\label{Eq-IntEnergy}
F_\text{int}(\bm{\upeta},\nabla \bm{\upeta})=\sum_{i=0}^N \sum_{j=i+1}^N \frac{4\gamma_{ij}}{\pi \ell_{ij}}\Big( \eta_i \eta_j - \ell_{ij}^2 \nabla \eta_i \cdot \nabla \eta_j \Big),
\end{equation}
where $\gamma_{ij}$ represents the interfacial energy density (per unit area) associated to the diffuse interface between phases $i$ and $j$, and $\ell_{ij}$ represents the corresponding interface thickness parameter such that the thickness of the diffuse interface between phases $i$ and $j$ in the direction normal to the interface and in a stress-free state is equal to $\lambda_{ij}=\pi \ell_{ij}$.

A viscous-type dissipation potential $D$ is included in the present model and is defined in terms of the rates of the order parameters $\dot{\bm{\upeta}}$ in the following form,
\begin{equation}\label{Eq-RateDiss}
D(\dot{\bm{\upeta}})=\sum_{i=0}^N \frac{\dot{\eta}_i^2}{2m_i},
\end{equation}
where $m_i$ denotes the mobility parameter and controls the propagation speed of the moving interfaces. It can be easily shown that the effective mobility parameter that governs propagation of the interface between phases $k$ and $l$ (when no other phases coexist) reads $m_{kl}=m_k m_l/(m_k+m_l)$ \cite{rezaee2020phase}.

\subsection{Incremental energy minimization framework}\label{Sec-energyMin}
We now formulate the evolution problem for the coupled phase-field equations by following the variational approach developed by Hildebrand and Miehe \cite{hildebrand2012phase}. In this approach, the complete evolution problem is formulated in a rate form as an (unconstrained) minimization problem, which is then consistently transformed into an incremental (time-discrete) problem, again in the form of a minimization problem.
Consideration of the physical inequality constraints on the order parameters leads to a constrained minimization problem \cite{tuuma2016size}, see also \cite{tuuma2018rate} for the case with a mixed viscous and rate-independent dissipation. Here, we skip the formulation of the rate evolution problem and directly introduce the problem in an incremental setting.

In the time-discrete setting, the solution $(\bm{\upvarphi}_n,\bm{\upeta}_n)$ at the previous time step $t_n$ is known and the fields $(\bm{\upvarphi}_{n+1},\bm{\upeta}_{n+1})$ at the current time step $t_{n+1}=t_n+\tau$ are sought, where $\tau>0$ is the time increment. The evolution problem is governed by the incremental energy minimization principle in which the incremental energy supplied to the system (which is equal to the increment of the potential energy and dissipation) is minimized. The global incremental potential is thus formulated as
\begin{equation}\label{Eq-GlobIncPot}
\Pi_\tau[\bm{\upvarphi},\bm{\upeta}]=\mathcal{E}[\bm{\upvarphi},\bm{\upeta}]-\mathcal{E}[\bm{\upvarphi}_n,\bm{\upeta}_n]+\mathcal{D}_\tau[\bm{\upeta}],
\end{equation}
where $\mathcal{E}$ and $D_\tau$ represent, respectively, the global potential energy functional and the global incremental dissipation potential, and the fields with no subscript, $(\bm{\upvarphi},\bm{\upeta})$, are those related to the current time step $t_{n+1}$. Note that the dependence of $\Pi_\tau$ on $(\bm{\upvarphi}_n,\bm{\upeta}_n)$ is not indicated, since these quantities are known.

The global potential energy functional $\mathcal{E}$ in Eq.~\eqref{Eq-GlobIncPot} is defined as the sum of the global Helmholtz free energy functional $\mathcal{F}$ and the potential of the external loads $\Omega$, thus
\begin{equation}
\mathcal{E}[\bm{\upvarphi},\bm{\upeta}]=\mathcal{F}[\bm{\upvarphi},\bm{\upeta}]+\Omega[\bm{\upvarphi}], \qquad \mathcal{F}=\int_B F(\nabla \bm{\upvarphi},\bm{\upeta},\nabla \bm{\upeta}) \rd V,
\end{equation}
with $B$ denoting the domain occupied by the body.
Note that in this study the external load is applied through the contact between the indenter and the body, thus $\Omega=0$, see Section \ref{Sec-contact}.

On the other hand, the global incremental dissipation potential $\mathcal{D}_\tau$ takes the form
\begin{equation}
\mathcal{D}_\tau[\bm{\upeta}]=\int_B D_\tau(\bm{\upeta}) \rd V, \qquad D_\tau(\bm{\upeta})=\tau D\Big( \frac{\bm{\upeta}-\bm{\upeta}_n}{\tau} \Big)=\sum_{i=0}^N \frac{\tau}{2m_i} \Big( \frac{\eta_i-\eta_{i,n}}{\tau} \Big)^2,
\end{equation}
where the local time-discrete dissipation potential $D_\tau$ is obtained by applying the backward Euler method to the local dissipation rate-potential $D$ in Eq.~\eqref{Eq-RateDiss}.

Finally, the evolution of the fields of $\bm{\upvarphi}$ and $\bm{\upeta}$ is obtained by the minimization of the constrained global incremental potential $\Pi_\tau^*$ as
\begin{equation}\label{Eq-EvolutionPr}
\{ \bm{\upvarphi},\bm{\upeta} \}= \arg \min_{\substack{\bm{\upvarphi},\bm{\upeta}}} \Pi_\tau^*[\bm{\upvarphi},\bm{\upeta}], \qquad \Pi_\tau^*[\bm{\upvarphi},\bm{\upeta}]=\Pi_\tau[\bm{\upvarphi},\bm{\upeta}]+\mathcal{I}_\mathcal{H}[\bm{\upeta}].
\end{equation}
The physical constraints on the order parameters $\eta_i$, cf.\ Eq.~\eqref{Eq-order}, are introduced in the global incremental potential $\Pi_\tau^*$ through the functional $\mathcal{I}_\mathcal{H}$,
\begin{equation}
\mathcal{I}_\mathcal{H}[\bm{\upeta}]=\int_B I_\mathcal{H}(\bm{\upeta}) \rd V, \qquad I_\mathcal{H}(\bm{\upeta})=\begin{cases}
0 & \bm{\upeta} \in \mathcal{H}, \\
+\infty & \text{otherwise},
\end{cases}
\end{equation}
where $I_\mathcal{H}$ is the indicator function of the admissible set (standard simplex) $\mathcal{H}$,
\begin{equation}
\mathcal{H}=\{ \bm{\upeta} \in \mathcal{R}^{N+1}: \; 0\leq \eta_i, \; \sum_{i=0}^N \eta_i=1 \}.
\end{equation}

\subsection{Contact formulation}\label{Sec-contact}
In the context of indentation problems, the external load in the present study is modeled through frictionless contact with a rigid indenter. A brief description of the contact problem is presented in this section, see \cite{wriggers2006computational} for a more general and detailed presentation.

The indenter is represented by a rigid surface denoted by $\bar{\Gamma}$.
On the other hand, a part of the boundary $\partial B$ (in the reference configuration) constitutes the potential contact surface $\Gamma_\text{c}$. The current position of each point on the contact surface $\Gamma_\text{c}$ is defined by the mapping $\bm{\upvarphi}$, namely $\bfm{x}=\bm{\upvarphi}(\bfm{X})$, where $\bfm{X} \in \Gamma_\text{c}$. Upon exploiting the closest-point projection method, a one-to-one relationship is established between the points on the contact surface $\Gamma_\text{c}$ and their correspondents on the rigid surface $\bar{\Gamma}$, the latter described by $\bar{\bfm{x}}$. This leads to the definition of the kinematic contact variable $g_\text{N}$, called the normal gap,
\begin{equation}\label{Eq-Ngap}
g_\text{N}=(\bfm{x}-\bar{\bfm{x}}) \cdot \bar{\bfm{n}},
\end{equation}
where $\bar{\bfm{n}}$ represents the unit normal to $\bar{\Gamma}$ at $\bar{\bfm{x}}$.

The unilateral contact conditions describe the complementarity relationship between the normal gap $g_\text{N}$ and the normal contact traction $t_\text{N}$, viz.\
\begin{equation}\label{Eq-unilat}
g_\text{N} \geq 0, \qquad t_\text{N} \leq 0, \qquad g_\text{N} t_\text{N}=0.
\end{equation}
Accordingly, the frictionless contact interaction is incorporated into the formulation by imposing the impenetrability condition \eqref{Eq-unilat}$_1$ in the minimization problem \eqref{Eq-EvolutionPr}. An indicator function $I_{\mathcal{R}^+}$ of the set of all non-negative real numbers is thus introduced,
\begin{equation}
I_{\mathcal{R}^+}(g_\text{N})=\begin{cases}
0 & g_\text{N} \geq 0, \\
+\infty & \text{otherwise},
\end{cases}
\qquad \mathcal{I}_\text{c}[\bm{\upvarphi}]=\int_{\Gamma_\text{c}}\mathcal{I}_{\mathcal{R}^+}(g_\text{N}(\bm{\upvarphi})) \rd S,
\end{equation}
where the dependence of the normal gap $g_\text{N}$ on the deformation mapping $\bm{\upvarphi}$ is through Eq.~\eqref{Eq-Ngap} and $\bfm{x}=\bm{\upvarphi}(\bfm{X})$. The evolution problem \eqref{Eq-EvolutionPr} is therefore reformulated as
\begin{equation}\label{Eq-ProblemConst}
\{ \bm{\upvarphi},\bm{\upeta} \}= \arg \min_{\substack{\bm{\upvarphi},\bm{\upeta}}} \Pi_\tau^{*,\text{c}}[\bm{\upvarphi},\bm{\upeta}], \quad \Pi_\tau^{*,\text{c}}[\bm{\upvarphi},\bm{\upeta}]=\Pi_\tau^*[\bm{\upvarphi},\bm{\upeta}]+\mathcal{I}_\text{c}[\bm{\upvarphi}].
\end{equation}

\section{Finite-element treatment}\label{Sec-FEM}
The finite-element discretization of the present model is implemented in the framework of the Firedrake finite-element environment \cite{rathgeber2016firedrake}. The tight coupling between Firedrake and the PETSc library \cite{balay2019petsc} offers the use of several linear solvers and preconditioners. In view of the large-scale computations aimed in the present study, the GMRES iterative solver \cite{saad1986gmres} has been employed, combined with a geometric multigrid preconditioner with point-block Jacobi relaxation \cite{Trottenberg2001}. The use of Lagrange multipliers would yield a problem with a saddle-point structure, and for simplicity of the iterative solver we instead employ a standard penalty regularization technique to enforce the inequality constraints on the order parameters, Eq.~\eqref{Eq-order}, and the impenetrability contact condition, Eq.~\eqref{Eq-unilat}$_1$, see the details of the penalty regularization in Section~\ref{Sec-Penalty}. Firedrake features symbolic code manipulation and automatic differentiation techniques, which facilitates a straightforward implementation of elastic Hencky strain $\bfm{H}^\text{e}$, cf.\ Eq.~\eqref{Eq-elasticEn}$_2$, by employing the Pad\'e approximation method \cite{baker1996pade}, see Section~\ref{Sec-HenckyApprox}. The weak form of the governing equations and the finite-element discretization are presented in Section~\ref{Sec-GovWeak}. The details of the solution procedure and the computer implementation are provided in Section~\ref{Sec-FEDisc}.

\subsection{Penalty regularization method}\label{Sec-Penalty}
A satisfactory performance of the penalty regularization method in combination with the double-obstacle potential, cf.\ Eq.~\eqref{Eq-IntEnergy}, has already been demonstrated for 2D problems \cite{rezaee2020phase}. This approach is adopted also in the present study, and its suitability for large-scale problems treated by the iterative multigrid solver is examined in Section \ref{Sec-CompParStudy}.

Upon exploiting the sum-to-unity constraint, the phase volume fraction of austenite $\eta_0$ can be treated as a dependent variable and is thus defined as a function of the other order parameters, viz.\
\begin{equation}
\eta_0=\eta_0(\hat{\bm{\upeta}})=1-\sum_{i=1}^N \eta_i, \qquad \hat{\bm{\upeta}}=(\eta_1,\dots,\eta_N),
\end{equation}
so that $\bm{\upeta}=\bm{\upeta}(\hat{\bm{\upeta}})$, where $\hat{\bm{\upeta}}$ denotes the condensed vector of the order parameters. As a result, $N$ independent variables are used to define the system with $N+1$ phases, which obviously leads to computational savings.

Consequently, the evolution problem \eqref{Eq-ProblemConst} can be written as an unconstrained minimization problem,
\begin{equation}
\{ \bm{\upvarphi},\hat{\bm{\upeta}}\}=\arg \min_{\substack{\bm{\upvarphi},\hat{\bm{\upeta}}}} \hat{\Pi}_\tau^\text{pen}[\bm{\upvarphi},\hat{\bm{\upeta}}],
\end{equation}
where $\hat{\Pi}_\tau^\text{pen}$ is the penalty regularized global incremental potential of the following form,
\begin{equation}\label{Eq-PenPot}
\hat{\Pi}_\tau^\text{pen}[\bm{\upvarphi},\hat{\bm{\upeta}}]=\Pi_\tau[\bm{\upvarphi},\bm{\upeta}(\bm{\hat{\upeta}})]+\int_B \sum_{i=0}^N \frac{1}{2} \epsilon_\eta \langle \eta_i \rangle_-^2 \rd V+\int_{\Gamma_\text{c}} \frac{1}{2} \epsilon_\text{N} \langle g_\text{N} \rangle_-^2 \rd S,
\end{equation}
with $\epsilon_\eta$ and $\epsilon_\text{N}$ as the penalty regularization parameters associated with the physical constraints on the order parameters and the contact constraint, respectively, and the angular bracket $\langle \cdot \rangle_-$ indicates the following operation,
\begin{equation}
\langle x \rangle_-=\begin{cases}
0 & \text{if}\; x\geq0, \\
x & \text{otherwise}.
\end{cases}
\end{equation}

\subsection{Approximation of elastic Hencky strain $\bfm{H}^\text{e}$}\label{Sec-HenckyApprox}
An important issue in the finite-element implementation of the presented phase-field model arises in the evaluation of the elastic Hencky strain $\bfm{H}^\text{e}$, Eq.~\eqref{Eq-elasticEn}$_2$, which involves computation of the matrix logarithm, as well as its first and second derivatives. In this context, the most widely used methods are based on the spectral decomposition, e.g.~\cite{ortiz2001computation} or series expansion, e.g.~\cite{de2001exact}. Both methods are subject to certain computational limitations, for instance, numerical difficulties in the vicinity of repeated eigenvalues or lack of accuracy in low-order Taylor approximations. Instead, we opt to employ the Pad\'e approximation method \cite{baker1996pade}, which leads to a high-accuracy explicit formula for the matrix logarithm. Accordingly, the derivatives of the matrix logarithm can be computed directly, e.g.\ by means of an automatic differentiation algorithm.

The applicability of Pad\'e approximants for the evaluation of the elastic Hencky strain in hyperelasticity has been addressed recently by Rezaee-Hajidehi et al.~\cite{RezaeeHajidehi2020Pade}. In particular, the combination of high accuracy and computational efficiency of Pad\'e approximants, notably the Pad\'e approximant of low-order (2,2), has been illustrated. On account of this, the Pad\'e approximant of order (2,2) has been chosen in the present study, which leads to the following approximation of $\bfm{H}^\text{e}$,
\begin{equation}\label{Eq-pade22}
\bfm{H}^\text{e} \approx\bfm{H}^\text{e}_{(2,2)}=\frac{3}{2}(({\bfm{C}^\text{e}})^2-\bfm{I})(({\bfm{C}^\text{e}})^2+4\bfm{C}^\text{e}+\bfm{I})^{-1}.
\end{equation}
To justify the choice, a comparative study is performed in Section \ref{Sec-CompParStudy}, where the effect of the Pad\'e approximant order on the computations is investigated. To this end, approximations of orders (1,1) and (3,3) are also considered, which give the following approximations of the elastic Hencky strain $\bfm{H}^\text{e}$, respectively,
\begin{equation}\label{Eq-pade11}
\bfm{H}^\text{e} \approx\bfm{H}^\text{e}_{(1,1)}=(\bfm{C}^\text{e}-\bfm{I})(\bfm{C}^\text{e}+\bfm{I})^{-1},
\end{equation}
\begin{equation}\label{Eq-pade33}
\bfm{H}^\text{e} \approx\bfm{H}^\text{e}_{(3,3)}=\frac{1}{6}(11(\bfm{C}^\text{e})^3+27(\bfm{C}^\text{e})^2-27\bfm{C}^\text{e}-11\bfm{I})((\bfm{C}^\text{e})^3+9(\bfm{C}^\text{e})^2+9\bfm{C}^\text{e}+\bfm{I})^{-1}.
\end{equation}
For a more detailed discussion, see \cite{RezaeeHajidehi2020Pade}.

\subsection{Governing equations and finite-element discretization}\label{Sec-GovWeak}
The stationarity of the incremental potential $\hat{\Pi}_\tau^\text{pen}[\bm{\upvarphi},\hat{\bm{\upeta}}]$ with respect to the fields of $\bm{\upvarphi}$ and $\hat{\bm{\upeta}}$ defines the weak form of the mechanical equilibrium (virtual work principle),
\begin{equation}\label{Eq-WeakMech}
0=\delta_{\bm{\upvarphi}}\hat{\Pi}_\tau^\text{pen}[\bm{\upvarphi},\hat{\bm{\upeta}}]=\int_B \bfm{P} \cdot \nabla \delta \bm{\upvarphi} \rd V +\int_{\Gamma_\text{c}} t_\text{N} \delta g_\text{N} \rd S \qquad \forall \, \delta \bm{\upvarphi},
\end{equation}
and the evolution equation for the order parameters $\hat{\bm{\upeta}}$,
\begin{multline}\label{Eq-WeakOrder}
0=\delta_{\hat{\bm{\upeta}}} \hat{\Pi}_\tau^\text{pen}[\bm{\upvarphi},\hat{\bm{\upeta}}]=\int_B \sum_{i=1}^N \Bigg( \Bigg( \frac{\partial F}{\partial \eta_i}-\frac{\partial F}{\partial \eta_0}+\frac{\partial D_\tau}{\partial \eta_i}-\frac{\partial D_\tau}{\partial \eta_0}+\mu_i - \mu_0 \Bigg)\delta \hat{\eta}_i \\ + \Bigg( \frac{\partial F_\text{int}}{\partial \nabla \eta_i} -\frac{\partial F_\text{int}}{\partial \nabla \eta_0} \Bigg) \cdot \nabla \delta \hat{\eta}_i \Bigg) \rd V \qquad \forall \, \delta \hat{\bm{\upeta}},
\end{multline}
where $\delta \bm{\upvarphi}$ and $\delta \hat{\bm{\upeta}}$ denote admissible variations, $\bfm{P}=\partial F/\partial \bfm{F}$ represents the first Piola--Kirchhoff stress tensor, the (nominal) normal contact traction $t_\text{N}$ is expressed as $t_\text{N}=\epsilon_\text{N} \langle g_\text{N} \rangle_-$, the variation of the normal gap $g_\text{N}$ reads $\delta g_\text{N}=\bar{\bfm{n}} \cdot \delta \bm{\upvarphi}$, and $\mu_i=\epsilon_\eta \langle \eta_i \rangle_-$ is the contribution resulting from the penalty regularization of the inequality constraints, $\eta_i \geq 0$. 

The local form of the time-discrete evolution equation \eqref{Eq-WeakOrder} has the usual form of the Ginzburg--Landau equation, $\dot{\hat{\bm{\upeta}}}=-\hat{\bfm{M}} \, \delta \hat{\mathcal{F}}/\delta \hat{\bm{\upeta}}$, where $\hat{\bfm{M}}$ represents the symmetric (positive-definite) mobility matrix and $\delta \hat{\mathcal{F}}/\delta \hat{\bm{\upeta}}$ is the functional derivative of the global penalty-regularized free energy $\hat{\mathcal{F}}$, see Remark~1 in \cite{rezaee2020phase}.

We perform the spatial discretization of the problem using the finite-element method. The finite-element approximation of the global fields $\bm{\upvarphi}$ and $\hat{\bm{\upeta}}$ is thus obtained as,
\begin{equation}\label{Eq-FEapprox}
\bm{\upvarphi}^h=\sum_i N_i^{(\bm{\upvarphi})}\bm{\upvarphi}_i, \qquad \hat{\bm{\upeta}}{}^h=\sum_i N_i^{(\bm{\upeta})} \hat{\bm{\upeta}}_i,
\end{equation}
where $\bm{\upvarphi}_i$ and $\hat{\bm{\upeta}}_i$ are the nodal quantities, while $N_i^{(\bm{\upvarphi})}$ and $N_i^{(\bm{\upeta})}$ are the corresponding basis functions. We discretize the domain $B$ by standard isoparametric 4-noded tetrahedral elements and thus use piecewise-linear basis functions for both $\bm{\upvarphi}$ and $\hat{\bm{\upeta}}$ so that $N_i^{(\bm{\upvarphi})}=N_i^{(\bm{\upeta})}$ in the present implementation.

Next, the weak form of the mechanical equilibrium \eqref{Eq-WeakMech} and of the evolution equation for the order parameters \eqref{Eq-WeakOrder} are written in a compact form, viz.
\begin{equation}\label{Eq-WeakMechCompact}
\mathcal{G}_{\bm{\upvarphi}}[\bm{\upvarphi},\delta \bm{\upvarphi};\hat{\bm{\upeta}}]=0 \qquad  \forall \, \delta \bm{\upvarphi},
\end{equation}
\begin{equation}\label{Eq-WeakOrderCompact}
\mathcal{G}_{\bm{\upeta}}[\hat{\bm{\upeta}},\delta \hat{\bm{\upeta}};\bm{\upvarphi}]=0 \qquad \forall \, \delta \hat{\bm{\eta}},
\end{equation}
where the variables preceded by a semicolon refer to the additional arguments resulting from the coupling. Accordingly, the discretized weak forms are obtained via substituting the approximate fields $\bm{\upvarphi}^h$ and $\hat{\bm{\upeta}}{}^h$, Eq.~\eqref{Eq-FEapprox}, into Eqs.~\eqref{Eq-WeakMechCompact} and \eqref{Eq-WeakOrderCompact}, 
\begin{equation}
\mathcal{G}_{\bm{\upvarphi}}^h(\bfm{U},\delta \bfm{U};\bfm{V})=\mathcal{G}_{\bm{\upvarphi}}[\bm{\upvarphi}^h,\delta \bm{\upvarphi}^h;\hat{\bm{\upeta}}{}^h]=0 \qquad \forall \, \delta \bfm{U},
\end{equation}
\begin{equation}
\mathcal{G}_{\bm{\upeta}}^h(\bfm{V},\delta \bfm{V};\bfm{U})=\mathcal{G}_{\bm{\upeta}}[\hat{\bm{\upeta}}{}^h,\delta \hat{\bm{\upeta}}{}^h;\bm{\upvarphi}^h]=0 \qquad \forall \, \delta \bfm{V},
\end{equation}
with $\bfm{U}$ and $\bfm{V}$ as, respectively, the global vectors of the nodal displacements $\bfm{u}_i=\bm{\upvarphi}_i-\bfm{X}_i$ and order parameters $\hat{\bm{\upeta}}_i$. Since $\mathcal{G}_{\bm{\upvarphi}}^h$ and $\mathcal{G}_{\bm{\upeta}}^h$ are linear in $\delta \bfm{U}$ and $\delta \bfm{V}$, respectively, we have
\begin{equation}
\mathcal{G}_{\bm{\upvarphi}}^h(\bfm{U},\delta \bfm{U};\bfm{V})=\bfm{R}_{\bm{\upvarphi}}(\bfm{U};\bfm{V}) \cdot \delta \bfm{U}=0 \qquad \forall \, \delta \bfm{U},
\end{equation}
\begin{equation}
\mathcal{G}_{\bm{\upeta}}^h(\bfm{V},\delta \bfm{V};\bfm{U})=\bfm{R}_{\bm{\upeta}}(\bfm{V};\bfm{U}) \cdot \delta \bfm{V}=0 \qquad \forall \, \delta \bfm{V},
\end{equation}
where $\bfm{R}_{\bm{\upvarphi}}$ and $\bfm{R}_{\bm{\upeta}}$ are the corresponding global residual vectors. This finally leads to the following set of coupled nonlinear algebraic equations,
\begin{equation}\label{Eq-DiscNonEq}
\bfm{R}_{\bm{\upvarphi}}(\bfm{U};\bfm{V})=\bm{0}, \qquad \bfm{R}_{\bm{\upeta}}(\bfm{V};\bfm{U})=\bm{0}.
\end{equation}
Transition from the continuum weak forms \eqref{Eq-WeakMech}--\eqref{Eq-WeakOrder} to the discrete residuum form \eqref{Eq-DiscNonEq} involves standard steps, including numerical quadrature and assembly of the respective element quantities, see e.g.\ \cite{korelc2016}.

Recall that the residual $\bfm{R}_{\bm{\upeta}}(\bfm{V};\bfm{U})=\bm{0}$ describes a transient problem of evolution of the order parameters, hence it additionally depends on $\bfm{V}_n$, which is known from the previous time step and this dependence is thus not indicated in $\bfm{R}_{\bm{\upeta}}$.

\subsection{Solution procedure and computer implementation}\label{Sec-FEDisc}

The set of coupled nonlinear equations \eqref{Eq-DiscNonEq} is solved in a monolithic way, i.e.\ simultaneously with respect to all unknowns, by using Newton's method. The global residuals, Eq.~\eqref{Eq-DiscNonEq}, are thus rephrased in the following form,
\begin{equation}\label{Eq-FinalRes}
\bar{\bfm{R}}(\bar{\bfm{U}})=\{ \bfm{R}_{\bm{\upvarphi}},\bfm{R}_{\bm{\upeta}} \}=\bm{0}, \qquad \bar{\bfm{U}}=\{ \bfm{U},\bfm{V} \}.
\end{equation}
At each Newton iteration~$i$, a large set of linear algebraic equations has to be solved for $\Delta \bar{\bfm{U}}^{i}$,
\begin{equation}\label{Eq-linAlgEq}
\frac{\partial \bar{\bfm{R}}}{\partial \bar{\bfm{U}}} \, \Delta \bar{\bfm{U}}^{i}=-\bar{\bfm{R}}(\bar{\bfm{U}}^i), \qquad \bar{\bfm{U}}^{i+1}=\bar{\bfm{U}}^{i}+\Delta \bar{\bfm{U}}^{i},
\end{equation}
where $\partial \bar{\bfm{R}}/\partial \bar{\bfm{U}}$ denotes the global tangent matrix. Solving this linear system is typically the most computationally expensive stage in an implicit computational scheme. Since the final goal is to converge to the solution of the nonlinear problem \eqref{Eq-FinalRes}, it is not required to solve the corresponding linear sub-problems \eqref{Eq-linAlgEq}$_1$ accurately up to the machine precision, as in the case of a direct solver. Hence, an iterative solver is used and the solution vector $\Delta \bar{\bfm{U}}^i$ is sought within a given precision. As a result, the consumption of CPU resources is reduced to a large extent.

The automatic differentiation (AD) algorithm of the Unified Form Language (UFL) \cite{alnaes2014ufl} is employed in order to obtain the global residual and the global tangent matrix. Practically, the incremental potential $\hat{\Pi}_\tau^\text{pen}$, cf.\ Eq.~\eqref{Eq-PenPot}, is coded by the user in UFL and AD is then used to derive the residual $\bar{\bfm{R}}$ (the first derivative of $\hat{\Pi}_\tau^\text{pen}$), cf.\ Eq.~\eqref{Eq-FinalRes}, and the tangent $\partial \bar{\bfm{R}}/\partial \bar{\bfm{U}}$ (the second derivative of $\hat{\Pi}_\tau^\text{pen}$), cf.\ Eq.~\eqref{Eq-linAlgEq}. In a compact notation, the discretized global incremental potential is expressed as
\begin{equation}
\bar{\Pi}_\tau(\bar{\bfm{U}})=\hat{\Pi}_\tau^h(\bfm{U},\bfm{V})=\hat{\Pi}_\tau^\text{pen}(\bm{\upvarphi}^h,\hat{\bm{\upeta}}{}^h),
\end{equation}
and the global residual vector and the global tangent matrix are obtained by differentiating $\bar{\Pi}_\tau$ with respect to the global nodal quantities $\bar{\bfm{U}}$ as
\begin{equation}
\bar{\bfm{R}}(\bar{\bfm{U}})=\frac{\partial \bar{\Pi}_\tau}{\partial \bar{\bfm{U}}}, \qquad \frac{\partial \bar{\bfm{R}}}{\partial \bar{\bfm{U}}}=\frac{\partial^2 \bar{\Pi}_\tau}{\partial \bar{\bfm{U}} \partial \bar{\bfm{U}}}.
\end{equation}
Note that these AD capabilities involve {symbolic} manipulations of the weak form of the equations, rather than manipulation of Fortran or C++ code, and therefore do not incur any inefficiencies associated with low-level AD techniques.

Firedrake is a finite-element environment that is closely integrated with the PETSc library \cite{balay2019petsc}, which enables the use of a wide variety of linear solvers and preconditioners and thus allows for the solution of \eqref{Eq-linAlgEq}$_1$ in an efficient manner. Since the time-discretized problem is elliptic, geometric multigrid methods \cite{Trottenberg2001} can be employed. Using the PETSc library, the selected solvers can be composed in a flexible way \cite{brown2012composable}. The outer Newton iterations are provided by the PETSc object SNES (Scalable Nonlinear Equation Solver), which applies Newton's method. The linear sub-problems are solved with GMRES \cite{saad1986gmres} (using PETSc object KSPGMRES) with geometric multigrid used as preconditioner (using PETSc object PCMG).

Firedrake has convenient facilities for implementing multigrid solvers. The prolongation and restriction operators that map information between different discretization levels are provided by Firedrake \cite{kirby2018solver}. The problem on each level is constructed by reassembly, rather than Galerkin projection. The last ingredient is the smoother operation on each level of discretization for which the point-block Jacobi iteration (using PETSc object PCPBJACOBI) has been used. This uses a block Jacobi iteration, where the $9 \times 9$ blocks are formed from all unknowns collocated at a given node. 
A dense LU factorization with partial pivoting is used to invert the blocks. 
This block relaxation reflects the strong local coupling among solution components in the relaxation; using a na\"ive Jacobi relaxation (solving $1 \times 1$ blocks for each degree of freedom) causes the solver to fail at the onset of phase transformation.

For solving the linear system \eqref{Eq-linAlgEq}, the so-called V-cycle has been used. It is based on the idea of defect-correction iteration, where a small fixed number of pre-smoothing iterations is first applied followed by computing a correction on the coarser level of discretization. This procedure is repeated until the coarsest level is solved with the MUMPS library for sparse LU decomposition \cite{amestoy2001mumps}. Finally, the corrections are prolongated to the finer levels and used to correct the solution, with additional post-smoothing iterations applied.

The finite-element simulations reported in the subsequent section are carried out on the high-performance clusters operated by the IT4Innovations National Supercomputing Center in Ostrava, Czech Republic, namely, the Barbora cluster (BullSequana XH2000) consisting of 200 computing nodes, where each node possesses two 18-core Intel Xeon Gold 6240 processors (2.60 GHz, 192 GB RAM) with InfiniBand HDR, connected in a fat tree topology, running Red Hat Enterprise Linux Server release 7 \cite{Barbora}.

In all the simulations, an adaptive time-stepping strategy is applied that is based on the number of Newton iterations needed in the previous time step to converge to the solution. If the number is lower than the prescribed desired value (7 iterations), then the current time step is increased, otherwise if the number is higher than the desired value, the current time step is decreased. If the number of Newton iterations exceeds the limit value (12 iterations), the step is rejected and the current time step is reduced by half.

\section{Nano-indentation of a CuAlNi single crystal}\label{Sec-Example}
The main aim of the numerical examples provided in this section is to demonstrate the modeling capabilities of the computational phase-field model presented above and to examine selected relevant computational aspects. The microstructure evolution in a CuAlNi single crystal is thus considered as a model problem. First, the description of the problem is given in Section~\ref{Sec-ProbDesc}. The results of the microstructure evolution related to the reference simulation are presented in Section~\ref{Sec-ResultsMicro}. A parametric study is carried out in Section~\ref{Sec-CompParStudy}, where detailed analyses regarding the effects of interface thickness parameter $\ell$, penalty regularization parameters $\epsilon_\eta$ and $\epsilon_\text{N}$, and the order of Pad\'e approximant of the elastic Hencky strain $\bfm{H}^\text{e}$ are presented. Finally, the weak scaling performance of the model and the effect of mesh resolution are investigated in Section~\ref{Sec-WeakScal}.

\subsection{Problem description}\label{Sec-ProbDesc}
The microstructure evolution in a pseudoelastic CuAlNi shape memory alloy during nano-indentation is investigated in this section. CuAlNi exhibits a cubic-to-orthorhombic $\beta_1 \rightarrow \gamma_1'$ martensitic transformation involving $N=6$ variants of martensite characterized by the following transformation stretch tensors given here in the orthonormal basis of the cubic austenite unit cell \cite{Bhattacharya2003},
\begin{equation}
\bfm{U}_{1,2}^\text{t} =
\begin{pmatrix}
\frac{\alpha+\gamma}{2} & 0 & \pm\frac{\alpha-\gamma}{2} \\
0 & \beta & 0 \\
\pm\frac{\alpha-\gamma}{2} & 0 & \frac{\alpha+\gamma}{2}
\end{pmatrix}, \quad 
\bfm{U}_{3,4}^\text{t} =
\begin{pmatrix}
\frac{\alpha+\gamma}{2} & \pm\frac{\alpha-\gamma}{2} & 0 \\
\pm\frac{\alpha-\gamma}{2} & \frac{\alpha+\gamma}{2} & 0 \\
0 & 0 & \beta
\end{pmatrix}, \quad
\bfm{U}_{5,6}^\text{t} =
\begin{pmatrix}
\beta & 0 & 0 \\
0 & \frac{\alpha+\gamma}{2} & \pm\frac{\alpha-\gamma}{2} \\
0 & \pm\frac{\alpha-\gamma}{2} & \frac{\alpha+\gamma}{2}
\end{pmatrix}
\end{equation}
where stretch parameters $\alpha=1.0619$, $\beta=0.9178$ and $\gamma=1.0230$ are calculated from the lattice parameters. 

A computational domain of the size $L\times L \times H=350 \times 350 \times 200$ nm$^3$ is considered. A hierarchy of finite-element meshes with three levels of uniform mesh refinements, leading to four levels of finite-element discretization, has been employed in the restriction and prolongation cycles of the geometric multigrid solver. On the coarsest level of discretization, a finite-element mesh of $24\,000$ tetrahedral elements is set, which leads to approximately $52\,000$ degrees of freedom consisting of 3 displacements and 6 order parameters at each node. On the finest level of discretization, a finite-element mesh with approximately $19$ million degrees of freedom is obtained.

\begin{figure}
\begin{tabular}{c c c}
\hspace*{-2.0cm}\includegraphics[width=0.49\textwidth]{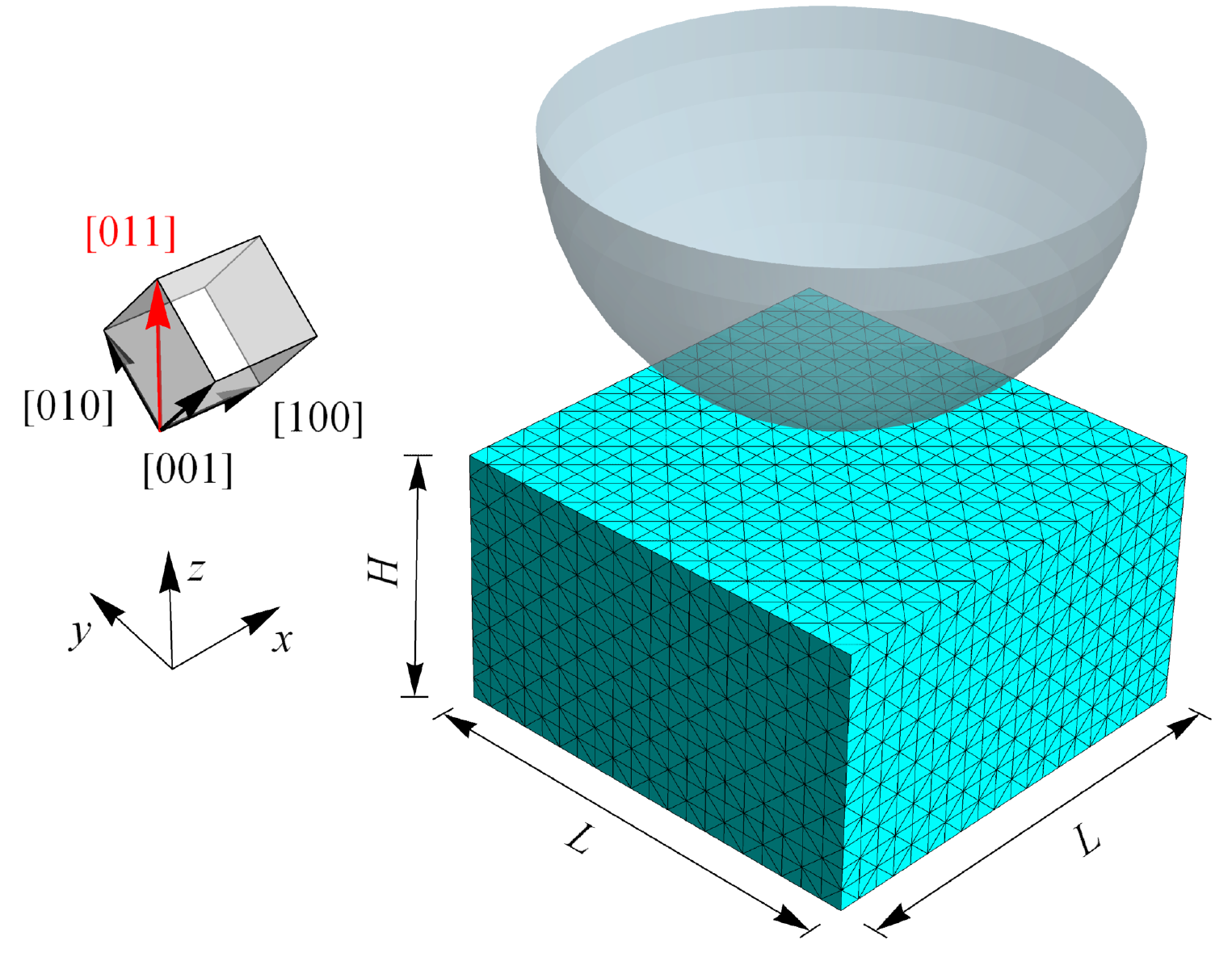} & \hspace*{0.3cm}\raisebox{-0.5cm}{\includegraphics[width=0.38\textwidth]{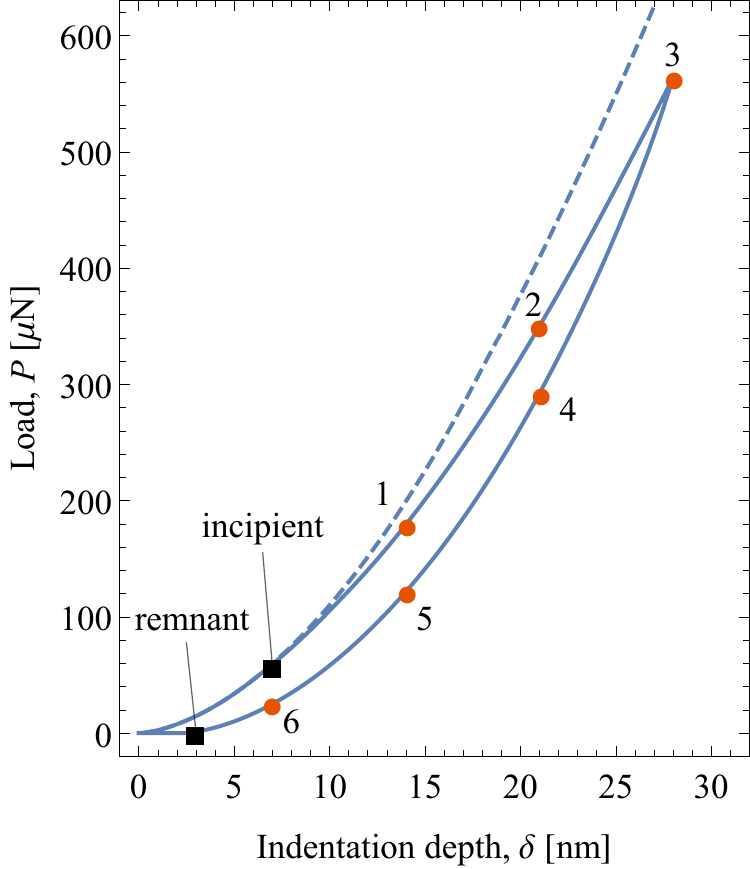}} & \hspace*{0.4cm} \raisebox{-0.6cm}{\includegraphics[width=0.27\textwidth]{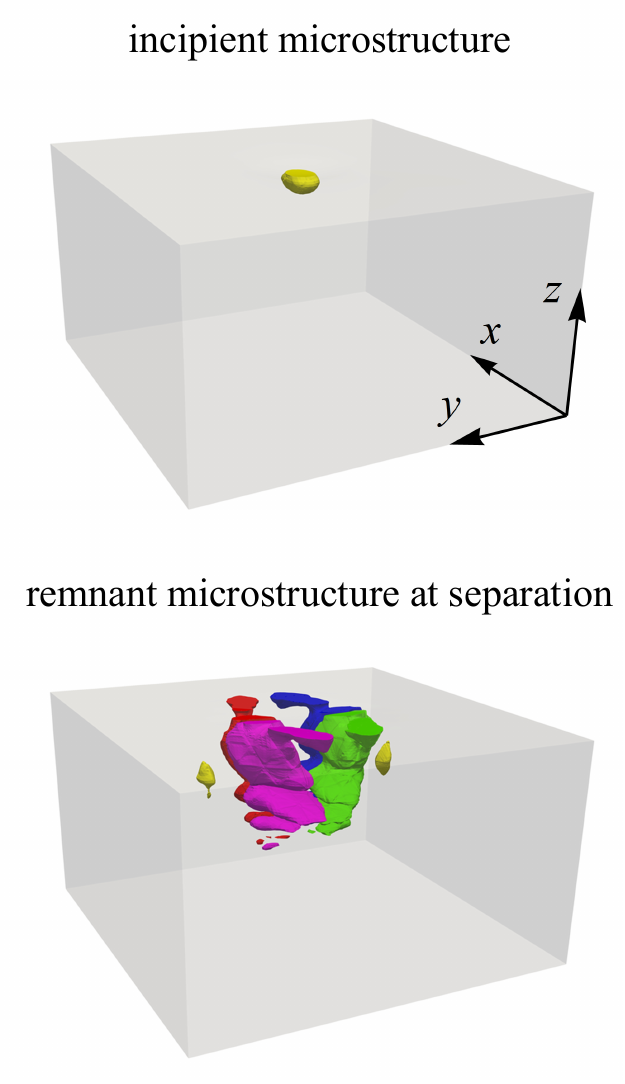}} \vspace*{0.2cm} \\
\hspace*{-2.2cm}\small{(a)} & \hspace*{0.3cm}\small{(b)} & \hspace*{0.4cm}\small{(c)}
\end{tabular}
\caption{Nano-indentation of a [011]-oriented CuAlNi single crystal: (a) the sketch of the indentation problem and the finite-element mesh (coarsest level), (b) the load--indentation depth ($P$--$\delta$) response, and (c) the incipient and remnant microstructures corresponding to the black markers in panel (b). The dashed line in panel (b) represents the elastic response and the numbered markers correspond to the snapshots in Fig.~\ref{Fig-snapshotAll}. To identify the martensite variants in panel (c), see the legend in Fig.~\ref{Fig-snapshotAll}.}
\label{Fig-geometry}
\end{figure}

A rigid spherical indenter of the radius $R=200$ nm is considered that compresses the block at the center of its top surface with a constant loading speed of $v=1$ nm/s. The loading continues up to the maximum indentation depth of $\delta_\text{max}=28$ nm, and afterwards, the indenter moves back to its initial position with the same speed. The vertical displacement of the bottom surface and the out-of-plane displacements of the lateral surfaces are constrained to be zero. The geometry of the problem and the finite-element mesh (on the coarsest level) are depicted in Fig.~\ref{Fig-geometry}(a).

It is assumed that the [011] axis of the austenite single crystal is parallel to the $z$ axis, i.e.\ the loading direction, see Fig.~\ref{Fig-geometry}(a). The elastic anisotropy of cubic austenite and orthorhombic martensite phases is taken into account and the anisotropic elastic constants are adopted from the literature data \cite{suezawa1976behaviour,YasunagaMeasurement}, see Table~\ref{tab-elConst}. The interfacial energy density for austenite--martensite interfaces $\gamma_{0i}=\gamma_\text{am}=0.2$ J/m$^2$ and martensite--martensite interfaces $\gamma_{ij}=\gamma_\text{mm}=0.02$ J/m$^2$ are adopted \cite{petryk2010interfacial}. The chemical energy of austenite is taken as  the reference, i.e.\ $F_0^0=F_\text{a}^0=0$, while those of martensite phases are taken as $F_i^0=F_\text{m}^0=5$ MPa, thus, implicitly, the temperature is defined in the pseudoelastic range such that the austenite is stable in stress-free conditions. The mobility parameters $m_i$ are the only time-dependent parameters of the present phase-field model. Here, $m_i=m=0.01$ (MPa s)$^{-1}$ are adopted so that reasonable predictions are provided by the present model for a physically relevant indentation speed, $v=1$ nm/s. Finally, the same interfcae thickness is considered for all interfaces, i.e.\ $\ell_{ij}=\ell=1$ nm, which results in $\lambda/h\approx 1.5$ on the finest level of discretization, where $\lambda=\pi \ell = 3.14$ nm is the theoretical interface thickness and $h$ denotes the element size. While the ratio $\lambda/h$ is here insufficient to resolve the diffuse interfaces with a high accuracy \cite{levitas2011phase}, the essential features of the solution are correctly captured, as shown in Section~\ref{Sec-WeakScal}.

\begin{table}[h]
\caption{Elastic constants of CuAlNi austenite and martensite single crystals \cite{suezawa1976behaviour,YasunagaMeasurement}. All in GPa.}
\vspace{1ex}
\label{tab-elConst}
\centering
\small{\begin{tabular}{lllllllllllll}
\hline
\multicolumn{3}{c}{Cubic $\beta_1$ phase}&~&\multicolumn{9}{c}{Orthorhombic $\gamma_1'$ phase} \\
\cline{1-3} \cline{5-13}
$c_{11}$&$c_{44}$&$c_{12}$&~&$c_{11}$&$c_{22}$&$c_{33}$&$c_{44}$&$c_{55}$&$c_{66}$&$c_{12}$&$c_{13}$&$c_{23}$\\
\hline
142&96&126&~&189&141&205&54.9&19.7&62.6&124&45.5&115\\
\hline
\end{tabular}
}
\end{table}

Based on preliminary computations, the penalty regularization parameters, $\epsilon_\eta=10^3$ GPa and $\epsilon_\text{N}=10^2$ GPa/nm, are chosen such that, on the one hand, the inequality constraints on the order parameters and the unilateral contact constraint are adequately enforced and, on the other hand, the performance of the computational scheme is not deteriorated by excessively large values of penalty parameters, see also Section \ref{Sec-CompParStudy}.

\subsection{Microstructure evolution}\label{Sec-ResultsMicro}
\begin{figure}
\begin{center}
\hspace*{-1.0cm}\includegraphics[width=1.11\textwidth,height=1.1\textheight,keepaspectratio]{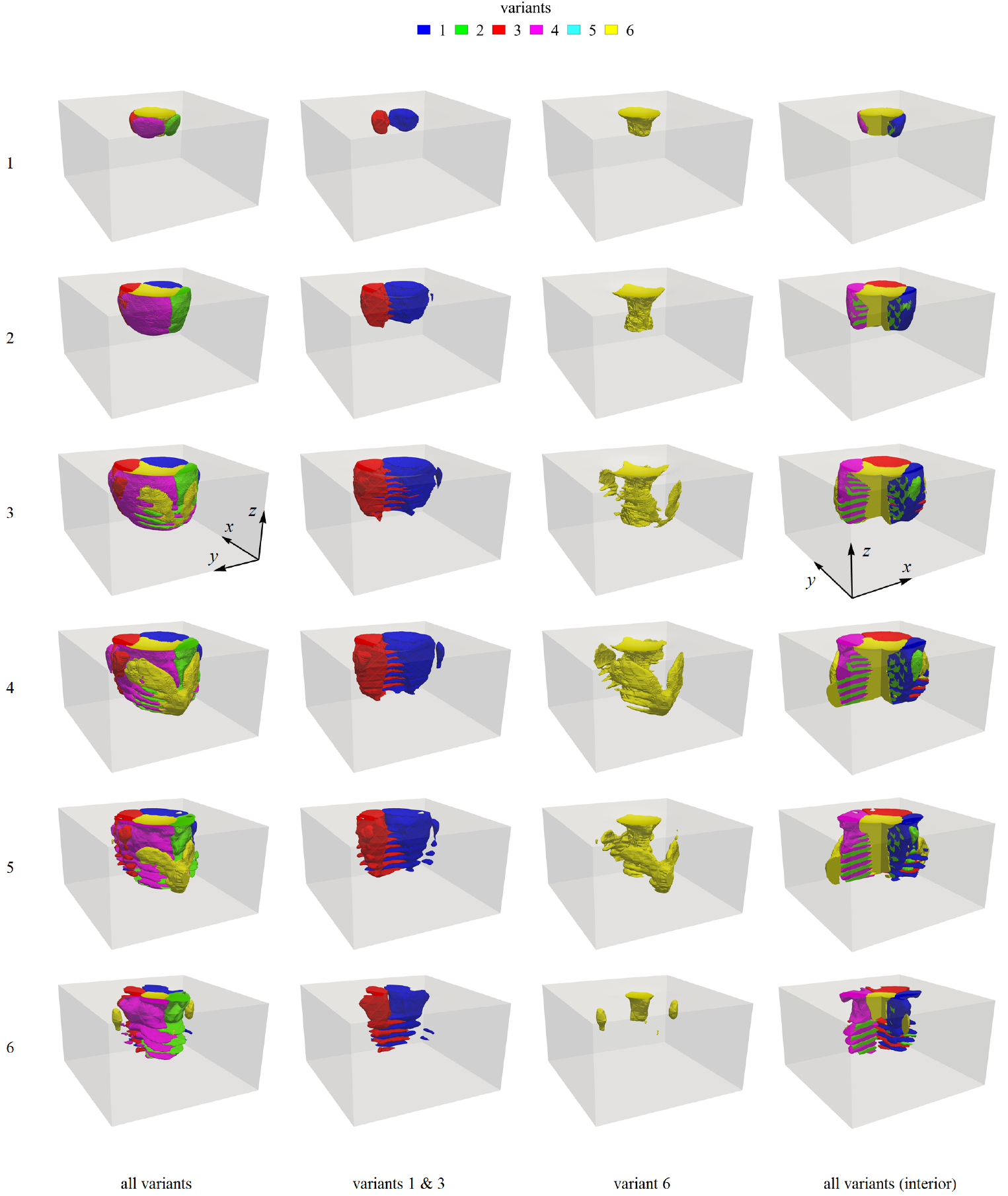}
\end{center}
\caption{Snapshots of the microstructure evolution in CuAlNi during nano-indentation at selected indentation depths, see the numbered red markers in Fig.~\ref{Fig-geometry}(b). Note that the interior snapshots are shown from a different angle as compared to the others, see the inset axes.}
\label{Fig-snapshotAll}
\end{figure}

Figure \ref{Fig-snapshotAll} depicts sets of selected snapshots of microstructure evolution during the loading--unloading process. The snapshots correspond to the numbered red markers superimposed on the load--indentation depth ($P$--$\delta$) curve in Fig.~\ref{Fig-geometry}(b), while the black markers in Fig.~\ref{Fig-geometry}(b) indicate the incipient and remnant microstructures, which are shown in Fig.~\ref{Fig-geometry}(c). The remnant microstructure stands for the transformation domain at the instant of separation of the indenter from the top surface during unloading. Each martensite variant $i$ is identified by a specific color and is represented by the domain of volume fraction $\eta_i \geq 0.5$. The domain of austenite and also the diffuse interfaces are excluded from the snapshots. The evolutions of the pair of martensite variants (1,3) and martensite variant 6 are illustrated separately in Fig.~\ref{Fig-snapshotAll}, so that a full picture of the microstructure evolution is provided. By symmetry, the former resembles the evolution of the pair (2,4). In addition, to examine the interior microstructure, one quarter of the transformed domain has been removed and the resulting snapshots are shown. Note that the snapshots of the interior microstructure have been rotated so that the most illustrative view is provided, see the inset axes in Fig.~\ref{Fig-snapshotAll}. 

The transformation initiates at an indentation depth of approximately $\delta=7$ nm by formation of an oval-shaped nucleus of martensite variant 6 below the indenter, see Fig.~\ref{Fig-geometry}(c), which constitutes the kernel of the transformation domain. The nucleation is not accompanied by an excursion event, namely a load drop or displacement burst (called pop-in), on the $P$--$\delta$ curve, as is often reported in nano-indentation experiments as an indication of incipient plasticity or phase transformation, e.g.~\cite{caer2013stress,laplanche2014sudden,dar2017nanoscale}. Our auxiliary simulation involving a slower loading process revealed a noticeable load drop on the $P$--$\delta$ curve at the initiation of the transformation. A clear transition from a sudden load drop to a sudden displacement burst, associated with the reduction of the stiffness of the indentation device, was observed in our previous 2D study \cite{rezaee2020phase}.

\begin{figure}
\begin{center}
\hspace*{-1.40cm}\includegraphics[width=1.1\textwidth,height=1.1\textheight,keepaspectratio]{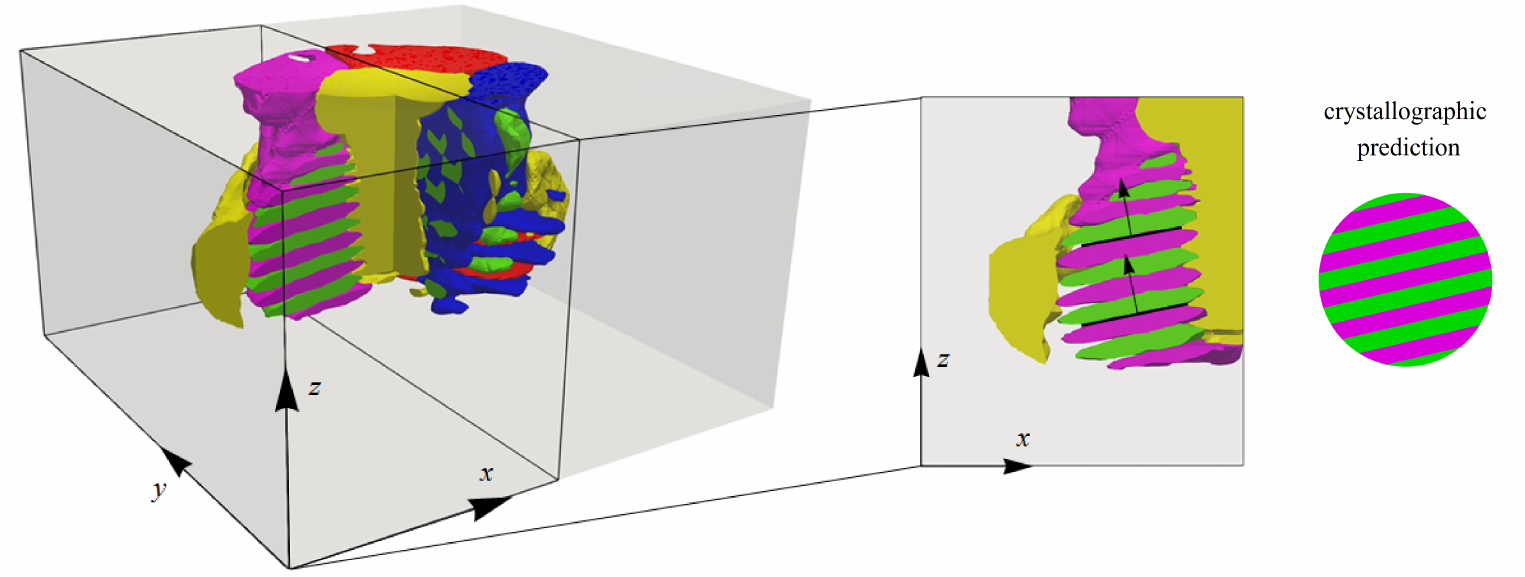}
\end{center}
\caption{The enlarged view of the interior microstructure (shown in the reference configuration) at $\delta=15$ nm during unloading, see the corresponding snapshot 5 in Fig.~\ref{Fig-snapshotAll}, providing a clear illustration of the saw-tooth and twinning morphologies. The stripped circle on the right presents the crystallographic theory prediction of the twin interface orientation (the interface normal lies within the ($x,z$)-plane).}
\label{Fig-enlargedinterior}
\end{figure}

At the indentation depth of approximately $\delta=10$ nm, four other martensite variants, the pairs (1,3) and (2,4), appear and surround the kernel of the transformation domain. The more the transformation domain grows, the more the $P$--$\delta$ curve deviates from the corresponding elastic curve. The microstructure starts developing a twinning pattern between the martensite pairs (1,3) and (2,4) at an indentation depth of about $\delta=20$ nm. Subsequently, the twinned martensite domains form a saw-tooth morphology with the kernel of the transformation domain through a zigzag-shaped interface layer. Similar patterns were obtained for austenite--twinned martensite interfaces by using a sharp-interface approach based on a shape-optimization technique \cite{stupkiewicz2007low} and the phase-field method \cite{tuuma2016size} for a two-dimensional periodic unit cell. A comparison has been made in Fig.~\ref{Fig-enlargedinterior} between the orientation of the twinning interfaces obtained for the martensite variant pair (2,4) in the present phase-field simulation and that predicted by the crystallographic theory \cite{Bhattacharya2003}. The comparison reveals a good agreement between computational and theoretical predictions, especially for those interfaces that are sufficiently far from the indenter, where the stress concentration is not high.

The transformation domain continues to grow at the early stage of unloading and starts to shrink with a delay. The delay is associated with the viscous evolution law, Eq.~\eqref{Eq-RateDiss}, see the related discussion in \cite{rezaee2020phase}.
The maximum size of the transformation domain is observed at an indentation depth of about $\delta=22$ nm during unloading. The reverse transformation proceeds with a fairly different evolution pattern as that observed during loading. The most notable differences are the persistence of the twinning patterns up to the final stage of unloading and the annihilation of the kernel of the transformation domain (variant 6) prior to the other variants. At $\delta=3$ nm, the indenter separates from the top surface, thus leading to zero external load $P=0$, while a remnant microstructure is still present, see the related snapshot in Fig.~\ref{Fig-geometry}(c), which disappears upon subsequent time progression. During the whole loading--unloading process, no sign of martensite variant 5 (with $\eta_5 \geq 0.5$) has been detected.

It is noteworthy that the microstructure evolution features symmetric and non-symmetric transformation modes. Initially, the transformation domain grows in a symmetric fashion, in agreement with the two-fold symmetry of the setup. However, the formation of the fine twins (see Fig.~\ref{Fig-enlargedinterior}) proceeds by an alternating development of the martensite variant plates and breaks the local symmetry of the microstructure, i.e.\ the symmetry that exists between the variants 1 and 3 and between the variants 2 and 4. At the same time, the overall symmetry of the microstructure is approximately maintained and remains unchanged during the whole process. A movie showing the complete microstructure evolution (movie M1) is available as the supplemental material accompanying this paper.

\subsection{Parametric study}\label{Sec-CompParStudy}
A parametric study is carried out in this section with the aim of investigating the effect of several important modeling parameters involved in the present phase-field model. This concerns the effect of the interface thickness parameter $\ell$, penalty regularization parameters $\epsilon_\eta$ and $\epsilon_N$, and Pad\'e approximation order, all being numerical parameters rather than physical. In fact, it is of primary importance to ascertain how these parameters influence the simulation results and the computational performance of the overall scheme in order to gain useful information regarding the functionality of the model and to ensure the reliability of the results.

To begin with, we discuss the effect of the interface thickness parameter $\ell$, cf.\ Eq.~\eqref{Eq-IntEnergy}. Additional simulations have been carried out for different values of $\ell$, namely $\ell=0.75,$ 2 and 4 nm, and the results are compared to those of our reference study (for $\ell=1$ nm). To keep a consistent rate of interface propagation as $\ell$ increases (decreases), the mobility parameter $m$ must decrease (increase) by the respective factor, see the related discussion in \cite{tuuma2018rate}. 
Therefore, $m=0.0133$, $0.005$ and $0.0025$ (MPa s)$^{-1}$ are adopted, respectively, for computations with $\ell=0.75$, 2 and 4 nm.
An expected effect of changing $\ell$ is reflected on the computational performance of the model. It has been seen that, as the value of $\ell$ increases, the computation proceeds faster, since the interfaces become more diffuse and can be resolved by the finite-element mesh more accurately. In contrast, for too low values of $\ell$, severe convergence issues are encountered, for instance, the simulation with $\ell=0.5$ nm was terminated before the end of the loading stage due to the failure of the Newton scheme.

\begin{figure}
\begin{tabular}{c c}
\hspace*{-2.5cm}\includegraphics[width=0.95\textwidth]{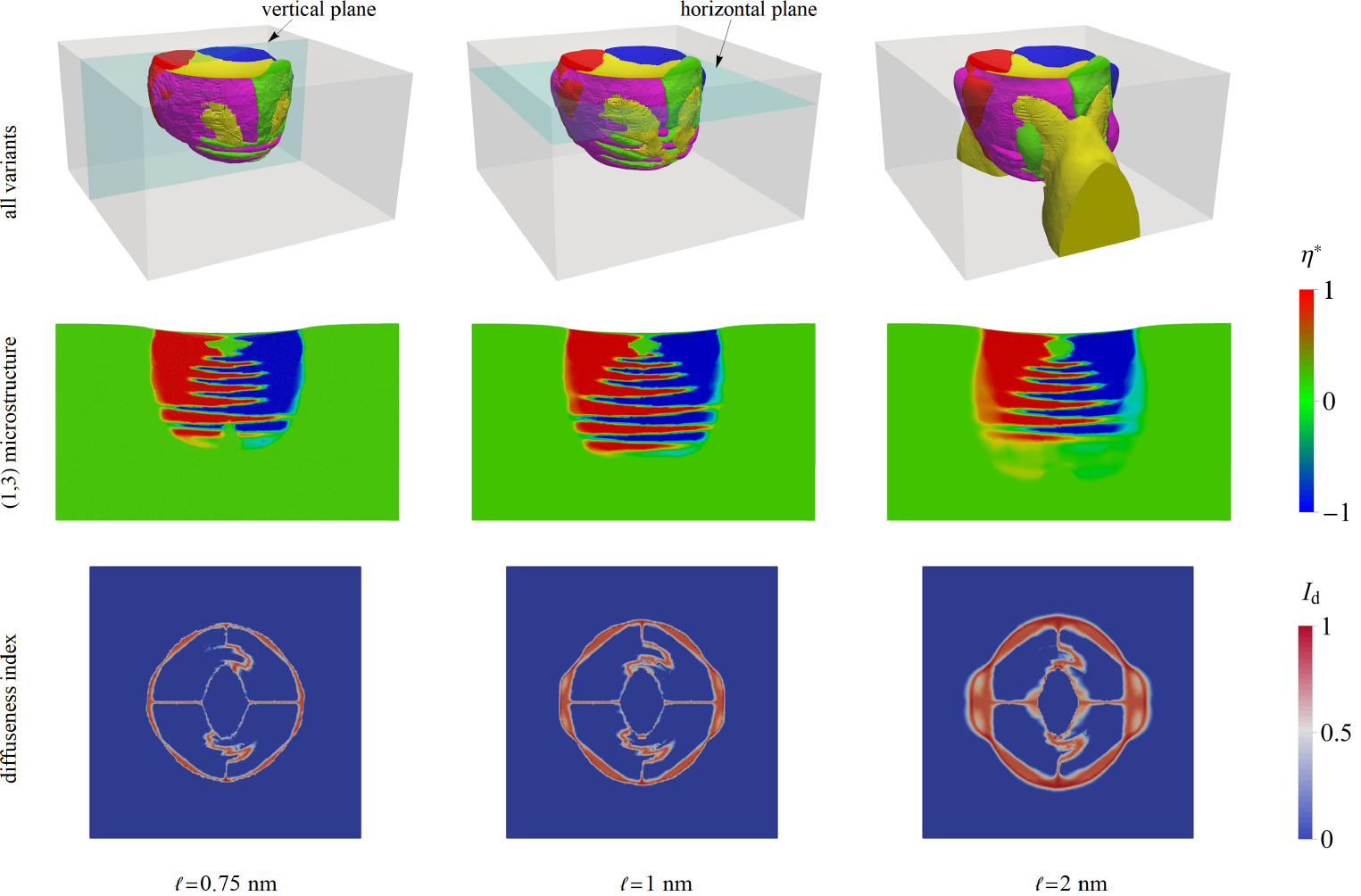}& \hspace*{0cm}\raisebox{1.5cm}{\includegraphics[width=0.34\textwidth]{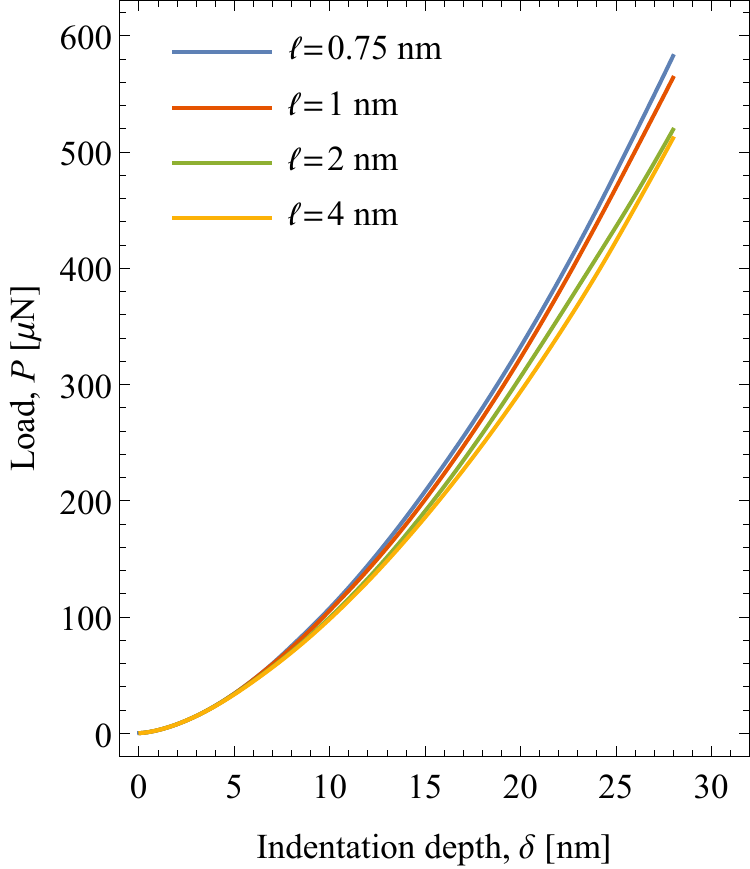}}\\
\hspace*{-3.3cm} \small{(a)} & \hspace*{0.3cm} \small{(b)}
\end{tabular}
\caption{The effect of the interface thickness parameter $\ell$ on (a) the microstructure and diffuseness of the interfaces at the end of loading, and (b) the $P$--$\delta$ response (loading only). The (1,3) microstructure in panel (a) is represented by $\eta^*=\eta_3-\eta_1$, and the diffuseness index by $I_\text{d}=1-\sum_{i=0}^{N} \eta_i^4$, see text.}
\label{Fig-effectLambda}
\end{figure}

Figure \ref{Fig-effectLambda} depicts the effect of parameter $\ell$ on the microstructure (at the end of loading) and on the $P$--$\delta$ response (loading only). As concerns the former, in addition to the effect of $\ell$ on the general pattern of the microstructure, the diffuseness of the interfaces has been also assessed quantitatively. 
To this end, the microstructure and the diffuseness of the interfaces are displayed at, respectively, a vertical plane (parallel to the $y$ axis and located 120 nm from the nearby lateral surface) and a horizontal plane (located 40 nm from the top surface). The microstructure is represented by a composite parameter $\eta^*=\eta_3-\eta_1$ so that $\eta^*=1$ corresponds to variant 3, $\eta^*=-1$ to variant 1 and $\eta^*=0$ to other phases (and also to the interface between variants 1 and 3). On the other hand, diffuseness of the interfaces is represented by the diffuseness index defined as $I_\text{d}=1-\sum_{i=0}^{N} \eta_i^4$, which takes the value of zero whenever any $\eta_i$ is equal to unity (pure phases) and values greater than zero within the diffuse interfaces.

Overall, the results show that, at a fixed indentation depth $\delta$, the size of the transformation domain increases with increasing $\ell$, and the load $P$ at the transformation branch of the $P$--$\delta$ response decreases. At the same time, the interfaces become obviously more diffuse. It can be seen that the general pattern of the microstructure, in terms of order of appearance and arrangement of the martensite variants and formation of the twinning patterns, is not affected by changing $\ell$. For the cases with large $\ell$, namely $\ell=2$ and 4 nm, separate domains of variant 6 appear at the bottom surface, which subsequently grow and join the main transformation domain, see the case of $\ell=2$ nm in Fig.~\ref{Fig-effectLambda}(a). 
This arises from the fact that, due to the computational restrictions, the size of the simulation domain is rather small, and spurious nuclei of variant 6 appear and develop at the boundary as a result of the interaction of the transformation domain with the boundary. Note also that the threshold of the driving forces associated with the transformation initiation decreases as $\ell$ increases, and thus the nucleation of the martensite variants occur at a lower stress.
For the sake of brevity, the simulation results for $\ell=4$ nm are not provided here. They present similar features as those for $\ell=2$ nm, except that the size of the transformation domain is larger and the interfaces are more diffuse. A detailed inspection of the microstructures reveals that, although some fine features of the microstructure observed in our reference simulation are still present for largely-diffuse microstructures, some details are missing. For instance, the saw-tooth pattern illustrated in Fig.~\ref{Fig-enlargedinterior} is hindered for $\ell=2$ and 4 nm.

Next, we report and discuss the effect of the penalty regularization parameters $\epsilon_\eta$ and $\epsilon_\text{N}$, which address, respectively, the inequality constraints on the order parameters, Eq.~\eqref{Eq-order}, and the contact constraint, Eq.~\eqref{Eq-unilat}$_1$. For this purpose, additional simulations are carried out for $\epsilon_\eta=200, 10^4$ and $10^5$ GPa and $\epsilon_\text{N}=10, 10^3$ and $10^4$ GPa/nm. Recall that, in our reference simulation, $\epsilon_\eta=10^3$ GPa and $\epsilon_\text{N}=10^2$ GPa/nm have been employed. 

\begin{figure}
\begin{tabular}{c c}
\hspace*{-2.5cm}\includegraphics[width=0.82\textwidth]{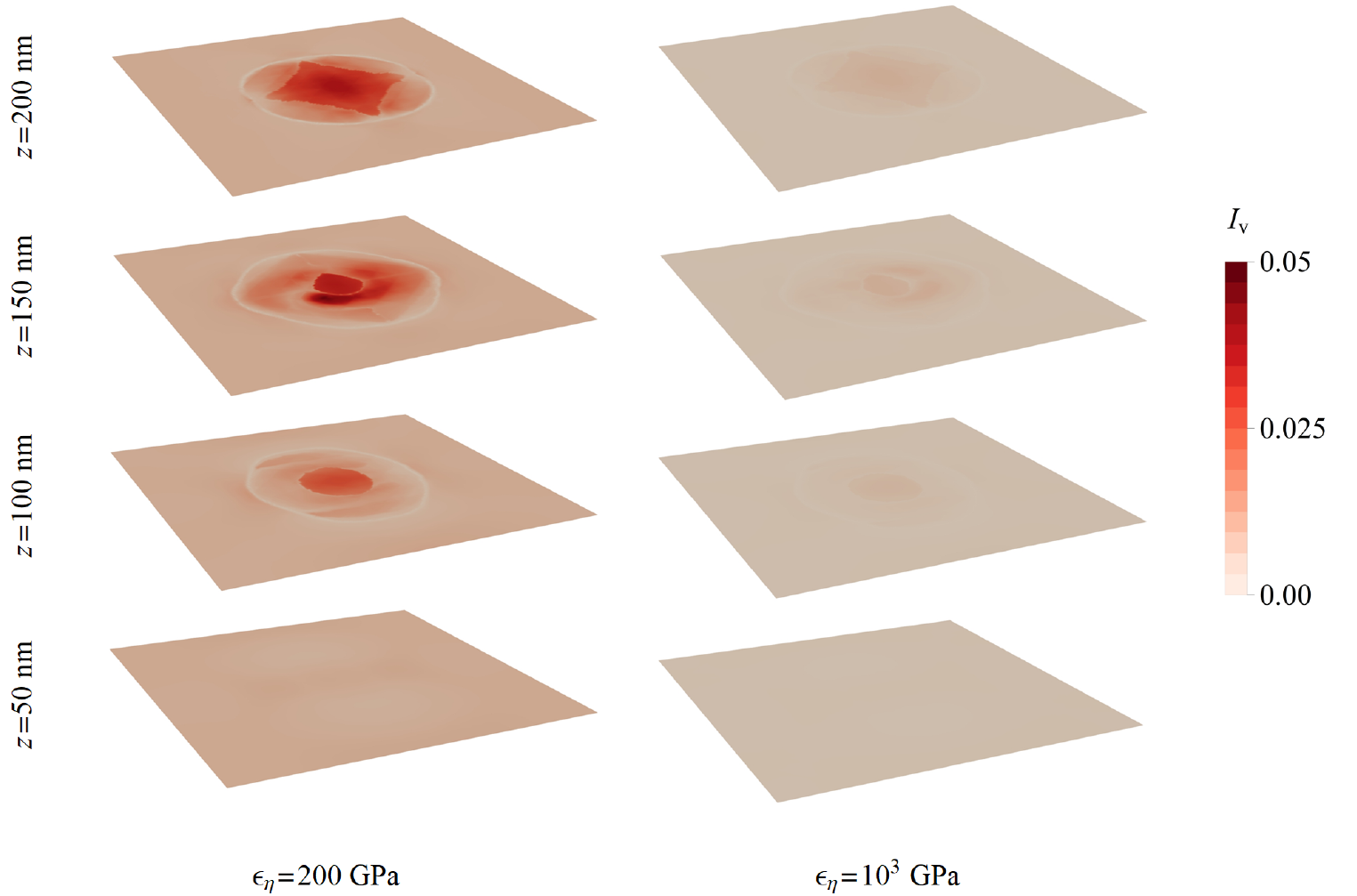}& \hspace*{0.3cm}\raisebox{1.0cm}{\includegraphics[width=0.42\textwidth]{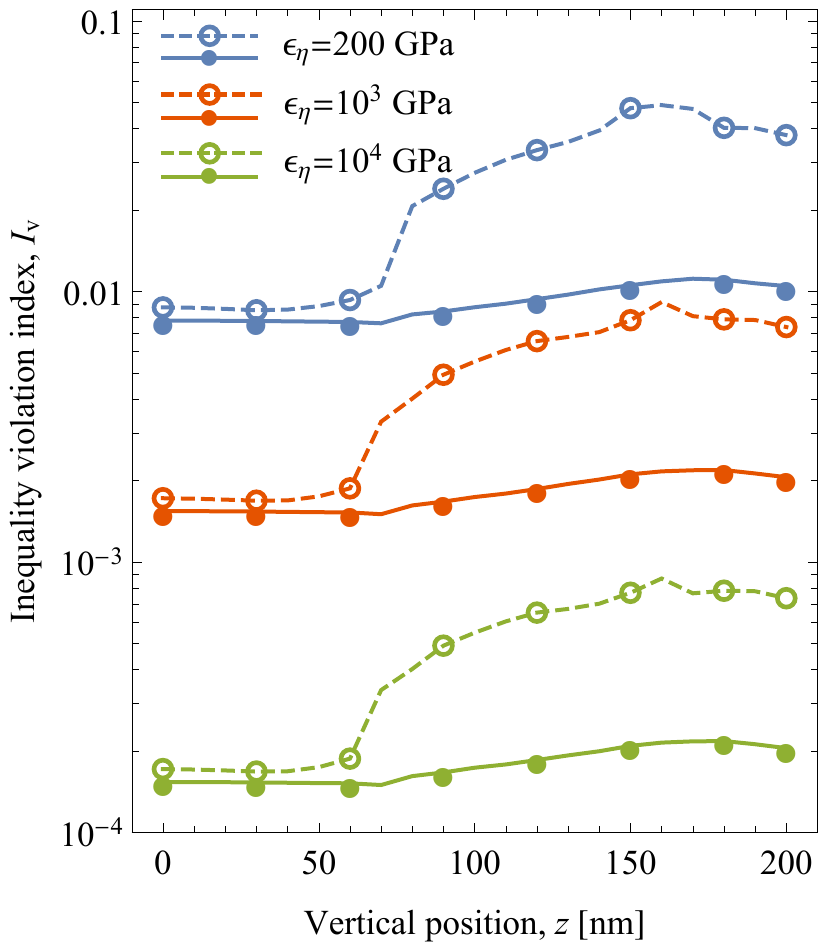}}\\
\hspace*{-2.5cm} \small{(a)} & \hspace*{0.3cm} \small{(b)}
\end{tabular}
\caption{The effect of the penalty parameter $\epsilon_\eta$ on the violation of the inequality constraints, $\eta_i \geq 0$, characterized by the inequality violation index $I_\text{v}=\sum_{i=0}^N \big| \langle \eta_i \rangle_- \big|$: (a) distribution of $I_\text{v}$ at selected horizontal planes and (b) the corresponding average (solid line) and maximum (dashed line) values of $I_\text{v}$ as a function of the position $z$ (taken in the reference configuration).}
\label{Fig-Iv}
\end{figure}

\begin{figure}
\begin{tabular}{c c c}
\hspace*{-2cm}\includegraphics[width=0.37\textwidth,height=1.1\textheight,keepaspectratio]{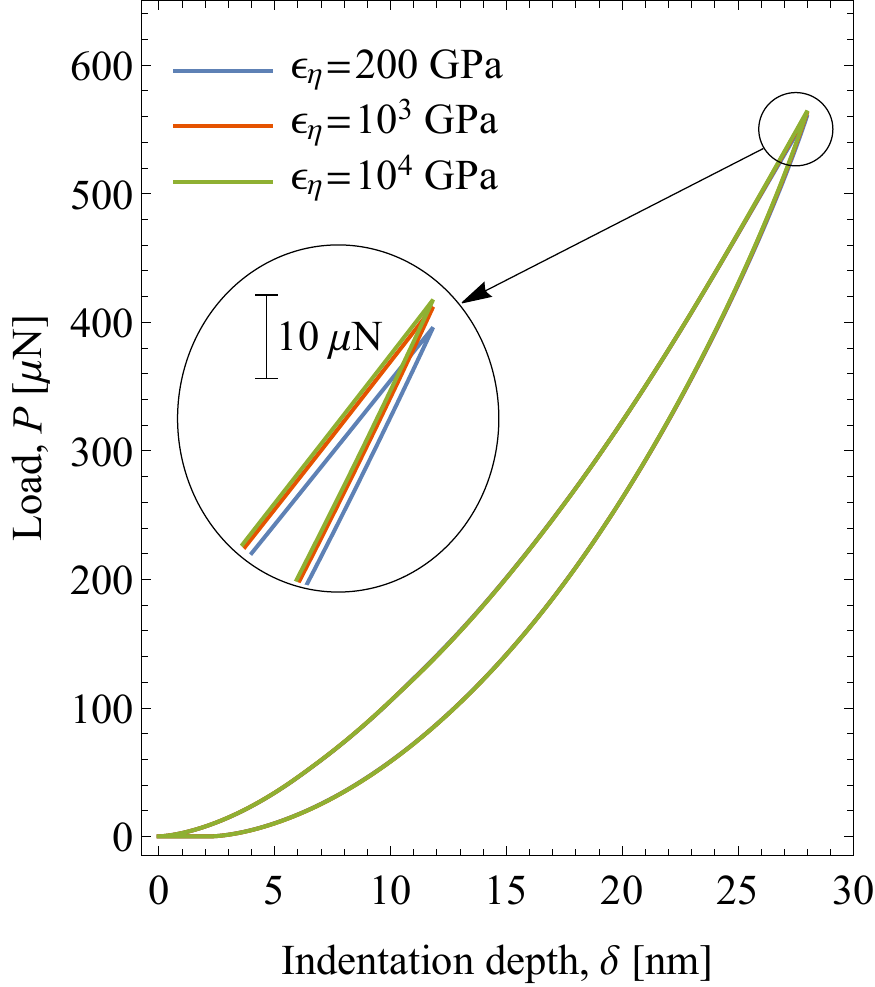} & \hspace*{0.2cm} \includegraphics[width=0.37\textwidth,height=1.1\textheight,keepaspectratio]{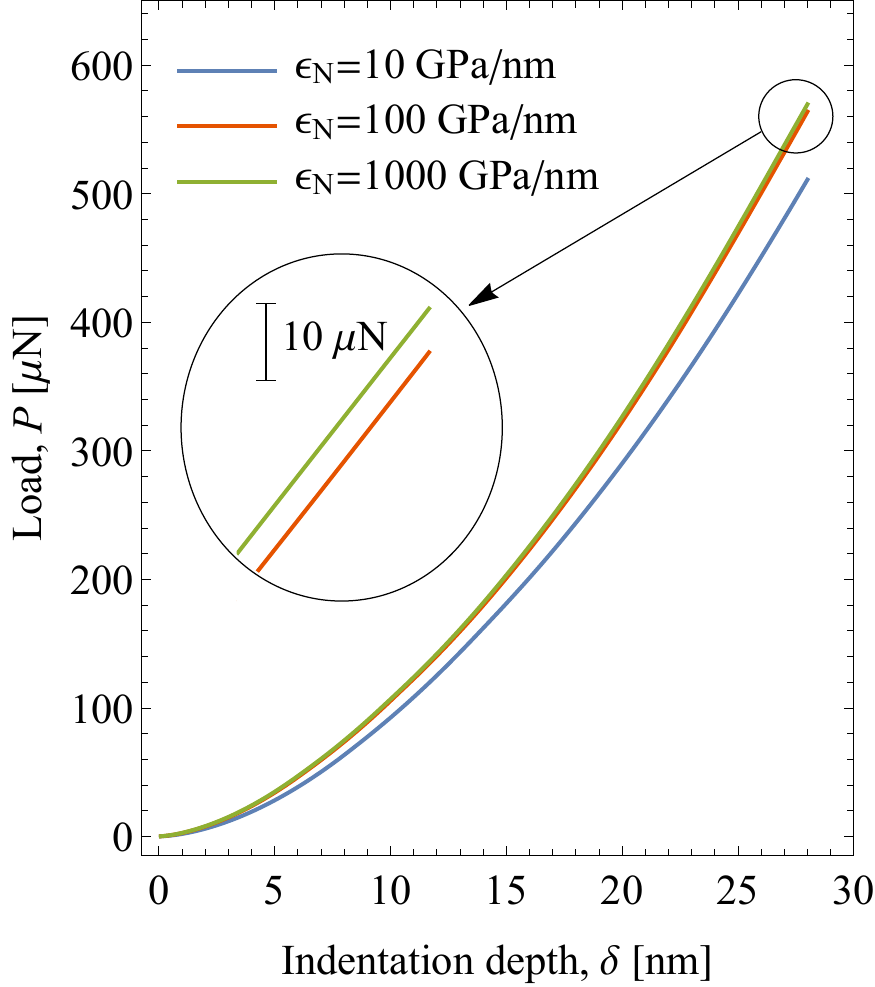} & \hspace*{0.2cm} \includegraphics[width=0.37\textwidth,height=1.1\textheight,keepaspectratio]{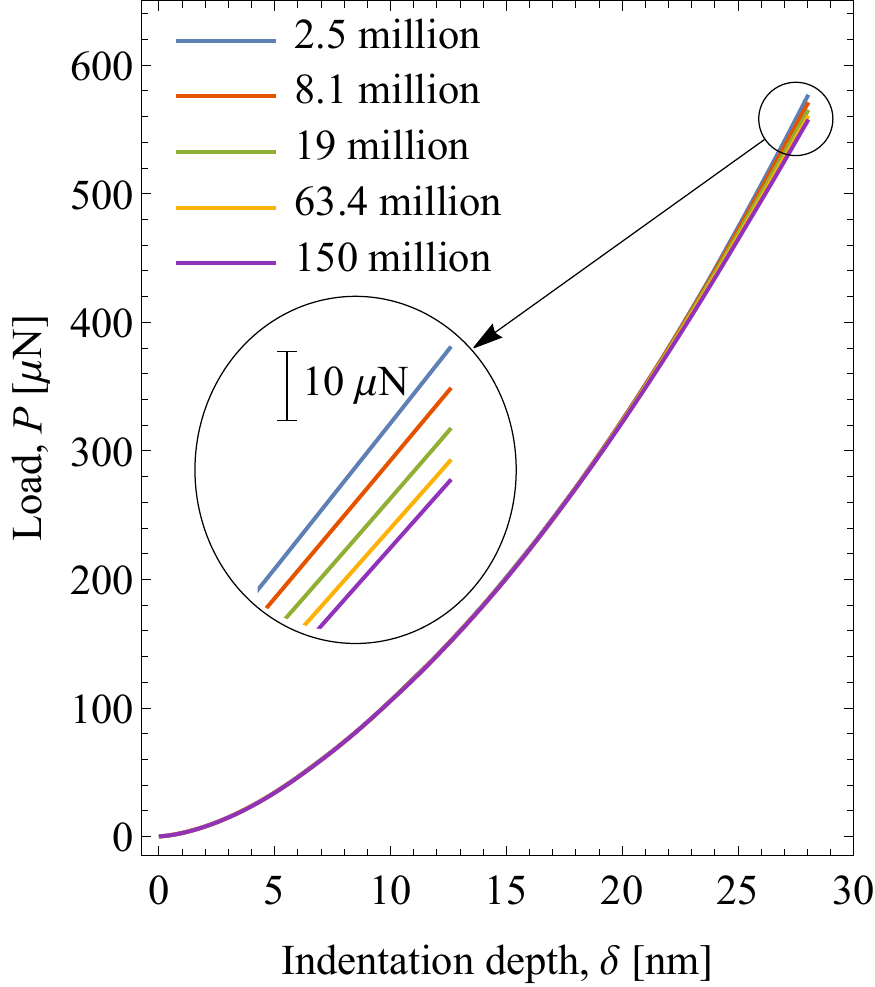} \\ 
\hspace*{-2cm} \small{(a)} & \hspace*{0.2cm} \small{(b)} & \hspace*{0.2cm} \small{(c)}
\end{tabular}
\caption{The effect of the penalty regularization parameters (a) $\epsilon_\eta$ and (b) $\epsilon_\text{N}$, and (c) the mesh resolution on the $P$--$\delta$ response. The legend in panel (c) refers to the number of degrees of freedom.}
\label{Fig-EpsEtaFD}
\end{figure}

In order to examine the violation of the inequality constraints $(\eta_i \geq0)$ as $\epsilon_\eta$ changes, a violation index $I_\text{v}=\sum_{i=0}^N \big| \langle \eta_i \rangle_- \big|$ is introduced, which provides a quantitative measure of the violation of all inequality constraints. Fig.~\ref{Fig-Iv} shows the distribution of the violation index $I_\text{v}$ at representative horizontal planes together with the graphs of the average and the maximum values of $I_\text{v}$ (calculated over the individual horizontal planes) as a function of the vertical position $z$. The first (and obvious) observation from Fig~\ref{Fig-Iv} is that the violation of the inequality constraints is reduced as the penalty parameter $\epsilon_\eta$ is increased. Secondly, the violation is more severe close to the indenter, which results from high stresses, and thus high driving forces for transformation beyond the physically admissible range $\eta_i \geq 0$. Noticeable violations are observed for $\epsilon_\eta=200$ GPa, with the maximum value of $I_\text{v}$ of about 0.05 at $z=150$ nm. For the reference simulation with $\epsilon_\eta=10^3$ GPa, the violations are barely visible in Fig.~\ref{Fig-Iv}(a) and the maximum value of $I_\text{v}$ is calculated as 0.007 at $z=150$ nm. For large values of $\epsilon_\eta$, namely $\epsilon_\eta=10^4$ and $10^5$ GPa, the violations are much smaller and no information can be extracted from the corresponding $I_\text{v}$ distribution plots (thus not provided here). The effect of $\epsilon_\eta$ on the $P$--$\delta$ response is shown in Fig.~\ref{Fig-EpsEtaFD}(a), revealing a negligible impact of $\epsilon_\eta$ even for $\epsilon_\eta=200$~GPa.

The present parametric study also encompasses the effect of the contact penalty parameter $\epsilon_\text{N}$ on the simulation results. Although such results are standard, a brief summary is provided here for completeness. The main conclusion is that the violation of the impenetrability contact constraint $g_\text{N} \geq 0$ is quite negligible for the reference simulation with $\epsilon_\text{N}=100$ GPa/nm and for those with higher $\epsilon_\text{N}$, while considerable violations have been detected for $\epsilon_\text{N}=10$ GPa/nm. The latter has led to a visible discrepancy in terms of the $P$--$\delta$ response, see Fig.~\ref{Fig-EpsEtaFD}(b).

To complete the picture, the computational performance of the model for different values of $\epsilon_\eta$ and $\epsilon_\text{N}$ is reported in Table \ref{Tab-PenReg}. In order to save the CPU resources, the computations for different contact penalty parameters $\epsilon_\text{N}$ are performed for loading only. It follows from Table \ref{Tab-PenReg} that within the range of the penalty parameters considered in this study, the efficiency of the computational scheme is not largely influenced by the choice of the penalty parameter, such that in both cases, the greatest simulation time (the sum of total linear solver time and total assembly time) is only about 1.4 times larger than the smallest one. It can be seen that the larger the penalty parameter $\epsilon_\eta$, the higher the number of Newton iterations, and thus the higher the number of time steps (recall that an adaptive time-stepping strategy is used, cf.\ Section \ref{Sec-FEDisc}). This emerges from the difficulty of solving the global nonlinear problem for higher $\epsilon_\eta$. At the same time, the decreasing trend of the average linear solver time per Newton iteration suggests that, in view of the smaller time steps, it becomes easier to solve the linear sub-problems. The situation is rather different for the contact penalty parameter $\epsilon_\text{N}$, and the case with the largest value, $\epsilon_\text{N}=10^4$ GPa/nm, exhibits an approximate doubling of the average linear solver time per Newton iteration (probably due to the deteriorated conditioning of the tangent matrix), which results in a longer simulation time. Table~\ref{Tab-PenReg} also reveals that, in the present implementation, the total assembly time constitutes the major contribution to the total simulation time. This is discussed further in Section \ref{Sec-WeakScal}.

\begin{table}
\caption{Computational performance of the phase-field model for various penalty regularization parameters $\epsilon_\eta$ (for loading and unloading) and $\epsilon_\text{N}$ (for loading only). The total assembly time refers to the time it takes to assemble the global residual vector and the global tangent matrix.}
\label{Tab-PenReg}
\vspace{1ex}
\centering
\small{
\begin{tabular}{llllllllll}
\hline
&\multicolumn{4}{c}{$\epsilon_\eta$ [GPa]}&~&\multicolumn{4}{c}{$\epsilon_\text{N}$ [GPa/nm]} \\
\cline{2-5} \cline{7-10}
&$200$&$10^3$&$10^4$&$10^5$&~&$10$&$10^2$&$10^3$&$10^4$ \\
\hline
Number of time steps&464&651&754&852&~&281&286&282&315\\
Total Number of Newton iterations $[10^{2}]$&39.9&40.8&48.5&57.4&~&14.8&16.1&15.3&16.7\\
Total number of linear solver iterations $[10^{3}]$&29.1&27.6&28.6&29&~&11.3&12.1&11.3&30.6\\
Total linear solver time $[10^3\, \text{s}]$& 13.5&13.3&14.2&15.1&~&5.6&5.7&5.6&13.3\\
Total assembly time $[10^3\, \text{s}]$&50.9&50.2&60.9&73.4&~&17.9&19.7&19&20.6\\
\hline
\end{tabular}
}
\end{table}

The last part of this section is focused on the effect of the Pad\'e approximant order of the elastic Hencky strain $\bfm{H}^\text{e}$. Following our recent study \cite{RezaeeHajidehi2020Pade}, a Pad\'e approximant of order (2,2), Eq.~\eqref{Eq-pade22}, has been employed to represent $\bfm{H}^\text{e}$, and it is important to determine whether and to what extent the simulation results are influenced by the choice of the Pad\'e approximant order. To this end, we repeated our reference simulation with Pad\'e approximants of order (1,1) and (3,3), see Eqs.~\eqref{Eq-pade11} and \eqref{Eq-pade33}. The corresponding data are listed in Table~\ref{Tab-pade}. It shows that, predictably, the total assembly time increases as the order of approximation increases, and the total assembly time for the model with the approximation order of (3,3) is nearly twice as long as that of the order (1,1). The response corresponding to the order (1,1) shows some small but noticeable deviation from the response obtained for the higher approximant orders (2,2) and (3,3). This, in particular, concerns the $P$--$\delta$ response at large indentation depths (the corresponding results are not provided for brevity). The responses corresponding to the approximant orders (2,2) and (3,3) are essentially identical, hence the choice of the order (2,2) is justified in view of its visibly lower computational cost.

\begin{table}
\caption{Computational performance of the phase-field model for various orders of the Pad\'e approximant used to approximate the elastic Hencky strain $\bfm{H}^\text{e}$ (for loading only).}
\label{Tab-pade}
\vspace{1ex}
\centering
\small{\begin{tabular}{llll}
\hline
Pad\'e approximant order &(1,1)&(2,2)&(3,3) \\
\hline
Number of time steps&282&286&282 \\
Total number of Newton iterations $[10^{2}]$&15.7&16.1&15.8 \\
Total number of linear solver iterations $[10^{3}]$&12.6&12.1&12 \\
Total linear solver time $[10^3\,\text{s}]$&6.1&5.7&5.7 \\
Total assembly time $[10^3\,\text{s}]$&15.2&19.7&30.5 \\
\hline
\end{tabular}
}
\end{table}

\subsection{Weak scaling performance}\label{Sec-WeakScal}
The results of the phase-field computations presented in the previous sections demonstrated several interesting and complex features of the microstructure. Capturing these features requires a fine and uniform finite-element mesh. On account of the fact that the model has 9 global degrees of freedom at each node\footnote{Note that for the numerical examples presented here, the phase-field model is specifically tailored for the study of phase transformation in pseudoelastic CuAlNi involving 6 variants of martensite, and thus involves 6 phase-field order parameters (in addition to 3 displacements) as degrees of freedom at each node. In the case of pseudoelastic NiTi, with a cubic-to-monoclinic transformation, 12 variants of martensite exist leading to a total of 15 degrees of freedom at each node, which would result in even computationally heavier simulations.}, such modeling becomes an intensive computational task, especially for 3D problems. The computational requirements can easily surpass the capabilities of a standard workstation and thus necessitate the use of parallel computing environments, such as clusters and supercomputers. It is therefore of great importance to test the parallel scaling performance of the present computational model. 

For this purpose, a weak scaling test has been carried out, where both the number of degrees of freedom of the problem and allocated CPU cores are scaled by the same factor. In this way, the single-core workload, i.e.\ the number of degrees of freedom per core, is kept approximately constant, at about 66\,000 degrees of freedom. Five different mesh resolutions are considered, corresponding to 2.5, 8.1, 19 (the reference simulation), 63.4 and 150 million degrees of freedom, which are solved on $1\times36=36$, ${\sim}3.4\times36=122$, $8\times36=288$, $27\times36=972$ and $64\times36=2304$ CPU cores respectively (out of a total of 7200 cores in the Barbora cluster, with 36 cores per node). 
Note that keeping the single-core workload at about 66\,000 degrees of freedom for the simulation with 8.1 million degrees of freedom would correspond to the allocation of 122 cores on 4 nodes (thus not all the 36 cores on each assigned node would be utilized). To make a meaningful analysis, the simulation with 8.1 million degrees of freedom has been run twice, once on $3\times36=108$ cores and once on $4\times36=144$ cores (thus using all cores on each assigned node), and the corresponding weak scaling data is determined by the linear (convex) combination of the data from the two simulations, in which the weights are chosen in such a way that the combination of the corresponding CPU cores is equal to 122.

Two families of finite-element meshes are employed in the present study: one consisting of $24\,000$ tetrahedral elements (approximately $52\,000$ d.o.f.) on the coarsest level of discretization and the other consisting of $10\,125$ tetrahedral elements (approximately $22\,000$ d.o.f.). The former has been employed for the mesh resolutions of 2.5, 19 and 150 million degrees of freedom by the use of, respectively, 2, 3 and 4 levels of uniform mesh refinements, whereas the latter has been employed for 8.1 and 63.4 million degrees of freedom by the use of, respectively, 3 and 4 levels of uniform mesh refinement.

\begin{figure}
\begin{tabular}{c c}
\hspace*{1.5cm} \includegraphics[width=0.35\textwidth]{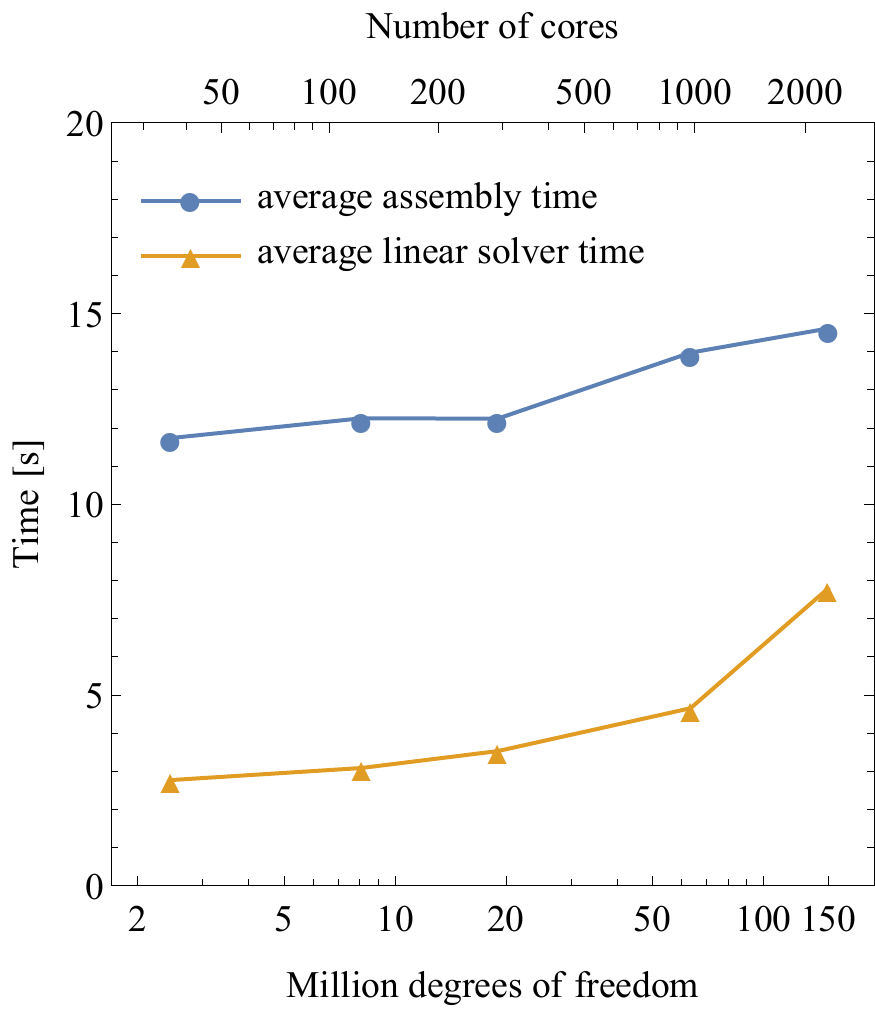}&\hspace*{0.5cm}\includegraphics[width=0.354\textwidth]{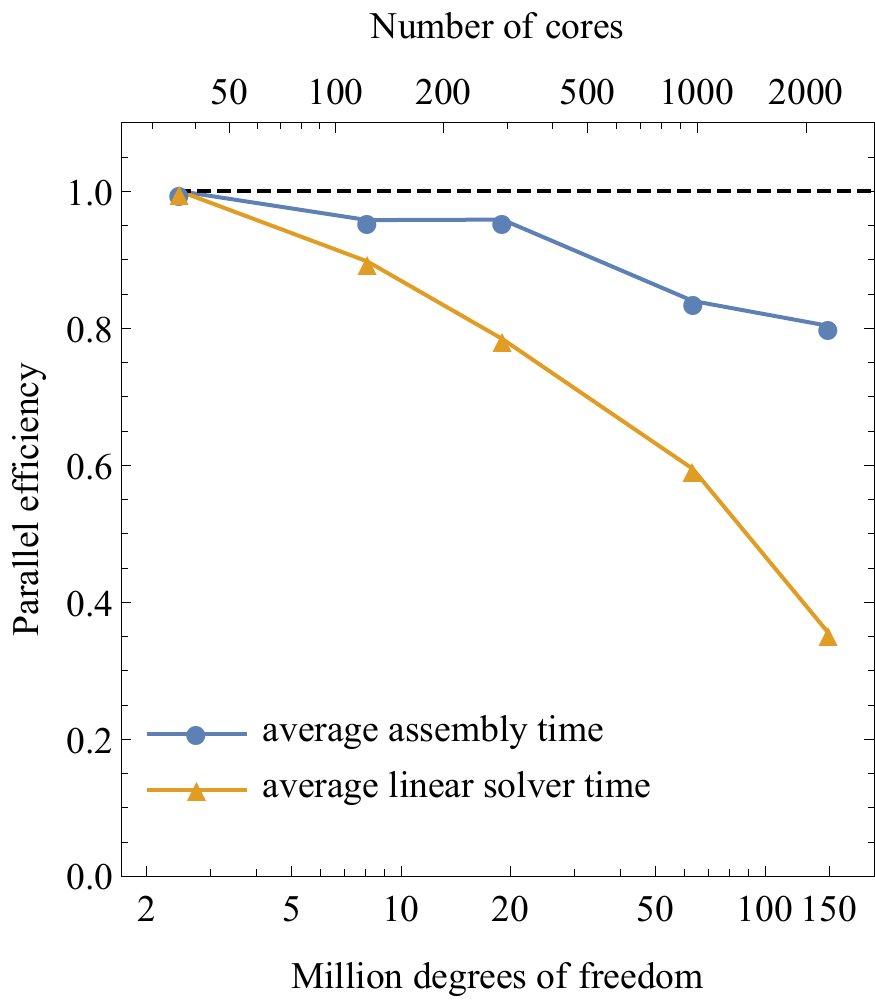}
\end{tabular}
\caption{Weak scaling performance: (a) average assembly and linear solver time per Newton iteration, and (b) the parallel efficiency as a function of problem size and number of allocated CPU cores. The dashed line in panel (b) indicates the ideal weak scaling efficiency. The corresponding detailed data are provided in Table \ref{Tab-weak}.}
\label{Fig-weakscal}
\end{figure}

The plots in Fig.~\ref{Fig-weakscal} present the weak scaling performance in terms of the average assembly and linear solver time per Newton iteration. The parallel efficiency reported in Fig.~\ref{Fig-weakscal}(b) is intuitively defined as the ratio between the average computational time for the simulation conducted on 36 cores (the smallest one with 2.5 million degrees of freedom) and that conducted on $36n$ cores. It follows that regarding the assembly time, which is more crucial compared to the linear solver time in the present implementation, a good weak scaling performance with a parallel efficiency of about $80\%$ is achieved for the largest simulation, which has led to an overall efficiency (assembly and linear solver) of about $65\%$. The detailed data of the weak scaling study is provided in Table~\ref{Tab-weak}. The data reveals that the number of time steps, and thus the total number of Newton iterations, is only marginally influenced by the mesh resolution, which reflects the robustness of the present implementation.

The weak scaling in terms of the linear solver time exhibits a rapid increase from the simulation with 63.4 million degrees of freedom to that of 150 million. It should be noted that the simulation with 150 million degrees of freedom has been run in two cluster racks, while the other simulations are performed within a single rack (in the Barbora cluster, each rack contains 40 nodes). Since the performance of the data transfer in an inter-rack node communication is lower compared to that in an intra-rack node communication, the computational performance of the finest simulation  is quite possibly degraded by a decrease in the communication performance as a result of the interaction of nodes residing in different racks. Note that, once beyond one rack, the communication performance will not be further affected by the number of racks involved in the computation. As a consequence, increasing the mesh resolution even further would not incur another rapid decrease in the communication performance.

\begin{table}
\caption{Details of the weak scaling performance: computational performance of the simulations with different problem size (for loading only).}
\label{Tab-weak}
\vspace{1ex}
\centering
\small{\begin{tabular}{llllll}
\hline
Number of assigned CPU cores (nodes)&36(1)&122$({\sim}3.4)$&288(8)&972(27)&2304(64) \\
Number of degrees of freedom $[10^{6}]$&2.5&8.1&19&63.4&150 \\
\hline
Number of time steps&282&281&286&286&297 \\
Total number of Newton iterations $[10^{2}]$&14.9&15.4&16.1&16.1&16.7 \\
Total number of linear solver iterations $[10^{3}]$&8.8&10.9&12.1&13.8&15.4 \\
Total linear solver time $[10^3\,\text{s}]$&4.1&4.7&5.7&7.5&13 \\
Total assembly time $[10^3\,\text{s}]$&17.5&18.8&19.7&22.5&24.4 \\
\hline
\end{tabular}
}
\end{table}

\begin{figure}[t]
\begin{center}
\hspace*{-1.8cm}\includegraphics[width=1.25\textwidth]{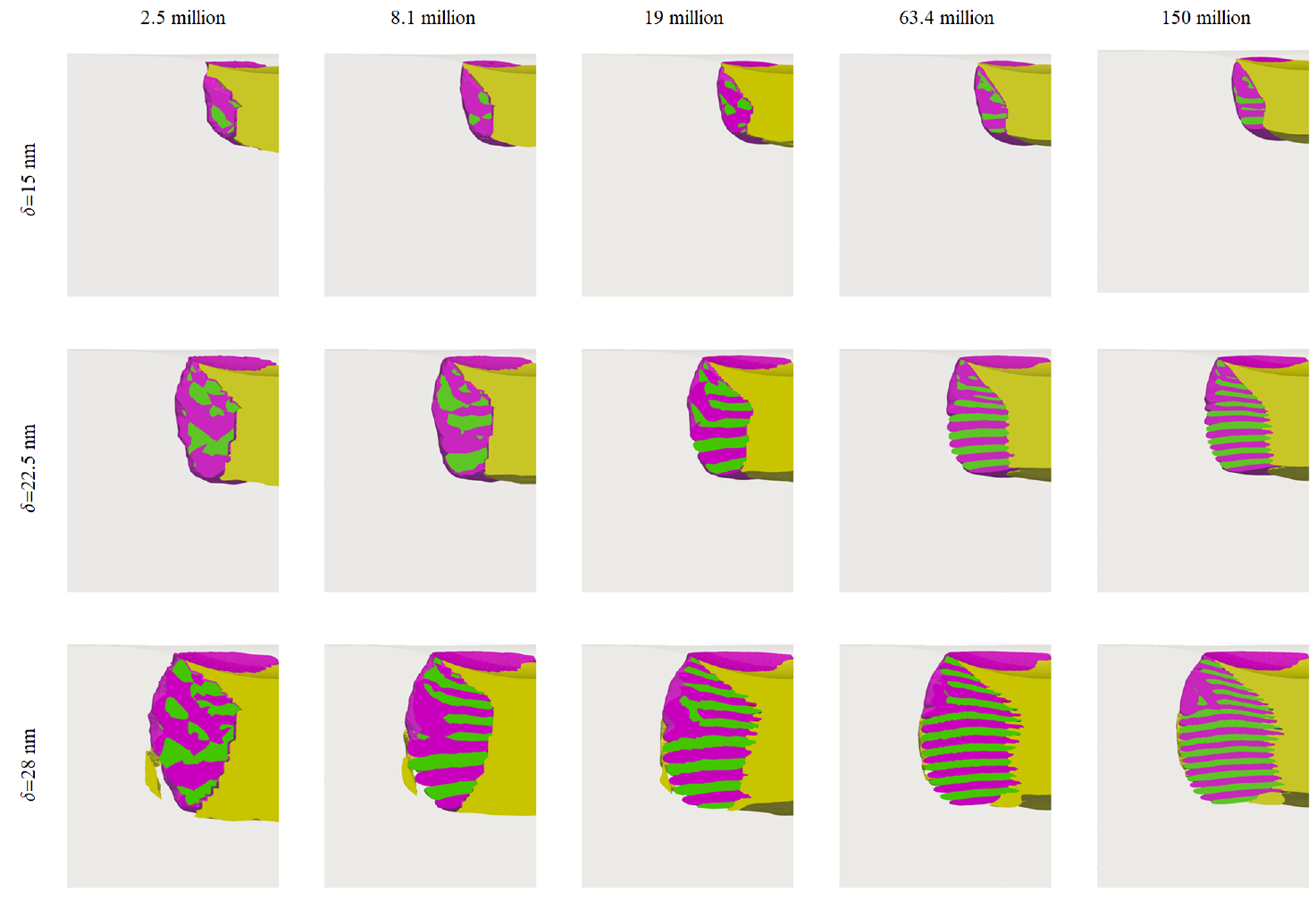}
\end{center}
\caption{The details of the microstructure (at selected indentation depths) for simulations with different mesh resolution. The numbers on top refer to the number of degrees of freedom.}
\label{Fig-meshmicro}
\end{figure}

Lastly, it is of interest to check how the simulation results, in particular the details of the microstructure and the $P$--$\delta$ response, are affected by the finite-element mesh resolution. In Fig.~\ref{Fig-meshmicro}, the details of the microstructure at selected indentation depths are illustrated for the five mesh resolutions. It is apparent that the general features of the microstructure are preserved in all cases. On the other hand, not all mesh resolutions have been capable of producing the fine features of the microstructure. For instance, the twinning pattern has not been properly captured for the lowest mesh resolution (2.5 million d.o.f.). Also, the saw-tooth pattern formed between the kernel of the transformation domain (variant 6) and the laminate of the martensite variant pair (1,3) is only visible for the cases with 19, 63.4 and 150 million degrees of freedom.
It can be seen that the twin spacing in the laminated region gradually decreases with the increase of the mesh resolution. It seems that a converged twin spacing has not been achieved yet, while a further mesh refinement cannot be afforded at this stage. According to the 2D study by Levitas and Javanbakht \cite{levitas2011phase}, at least 4--5 elements per (theoretical) interface width are needed to correctly resolve diffuse interfaces, and the interfacial energy is artificially increased if the mesh is not fine enough. This effect may explain why the microstructure gets finer as the mesh is refined, since the ratio $\lambda/h$ is here equal to approximately 1.5 in the reference case and to 3 in the case of the finest mesh. A movie showing the microstructure evolution (during loading) for different mesh resolutions (movie M2) is available as supplemental material accompanying this paper.

The effect of the mesh resolution on the $P$--$\delta$ response has been also examined, see Fig.~\ref{Fig-EpsEtaFD}(c). For the lowest resolution, the deviation of the load $P$ at the maximum indentation depth, with respect to that of the highest resolution, is approximately $20$~$\mu$N. As the mesh resolution increases, the deviation decreases, such that a deviation of only about 3 $\mu$N has been obtained for the case with $63.4$ million degrees of freedom.

\section{Conclusion}
A robust and scalable computational model has been developed for the simulation of multivariant martensitic phase transformations in shape memory alloys. The computational model combines an advanced finite-strain phase-field model and its finite-element discretization in Firedrake, including an iterative solver with multigrid preconditioner.

A reasonably good parallel scaling performance of the computational model has been demonstrated, such that the simulation of a complex high-resolution martensitic microstructure with the problem size reaching 150 million degrees of freedom has been successfully completed. Successful simulations have been carried out for a large range of model parameters and mesh resolutions, while at the same time the qualitative characteristics of the solution are preserved and the computational performance is not considerably affected, thus indicating the reliability and the robustness of the present model.

The indentation-induced microstructure evolution in a pseudoelastic CuAlNi shape memory alloy has been studied as an application of the present computational model. The predicted microstructure pattern revealed characteristic features, such as the formation of twinning microstructure and the subsequent development of the saw-tooth morphology. It has been observed that, upon refining the mesh resolution, although more refined microstructure patterns are revealed, the essential features of the microstructure are not affected. It follows that problems with larger, more physically relevant domain sizes can be modeled using a sufficiently fine mesh, in view of the trade-off between the fineness of the solution and the physical size of the problem.

It has been observed that, even if the overall two-fold symmetry of the microstructure is preserved during the whole process, the formation of the fine twins proceeds in a non-symmetric manner, and thus the symmetry is locally broken. Since the problem setup is symmetric, it might be tempting at first sight to solve the problem for only one quarter of the simulation domain (with proper symmetry conditions applied). However, it is evident from the present results that such an analysis would lead to an incorrect microstructure evolution.

A parametric study has been performed with the aim of examining the effect of selected numerical parameters. One of the notable outcomes of the parametric study concerns the performance of the penalty regularization method, which constitutes one of the key components of the present computational treatment of the double-obstacle potential. It has been observed that, within the wide range of the penalty parameters considered, the computational performance is not visibly affected, while the error introduced by the penalty regularization is insignificant. A satisfactory performance of the penalty method has thus been demonstrated, in particular, in the context of the iterative multigrid solver.

An issue that deserves further investigation is that the computational cost of the assembly is markedly larger than that of the linear solver, which is not usually the case in implicit finite-element schemes. The poor assembly performance is also reflected in the parametric study of the Pad\'e approximant order, where an increase in the order of approximation has led to a visible impact on the total assembly time. Although the issue could be partly attributed to the complexity of the phase-field model, our experience with other finite-element modeling environments (specifically AceGen/AceFEM \cite{korelc2016}, see also \cite{RezaeeHajidehi2020Pade}) indicates that there are substantial opportunities for the improvement of the assembly performance, and these will be pursued in the future.

Concluding, we note that the present simulations of the nano-indentation problem have been carried out for realistic and physically meaningful material parameters (elastic constants, transformation stretches and interfacial energies). The adopted indenter radius $R=200$ nm and the simulation domain size are relatively small. However, our study has demonstrated that the computational model is robust and scalable, hence with an appropriate supercomputer simulations can be readily carried out for larger and more physically relevant simulation domains.

\appendix

\section{Supplementary data}
Supplementary material related to this article can be found online at \url{https://doi.org/10.1016/j.cma.2021.113705}.

\paragraph{Acknowledgement} J.H. and K.T.\ have been supported by the Charles University Research program No.\ UNCE/SCI/023. K.T. has been supported by the Czech Science Foundation through the project 18-12719S. M.R.H. and S.S. have been supported by the National Science Center (NCN) in Poland through Grant No.\ 2018/29/B/ST8/00729. P.E.F.\ has been supported by EPSRC grants EP/R029423/1 and EP/V001493/1. This work was supported by the Ministry of Education, Youth and Sports of the Czech Republic from the Large Infrastructures for Research, Experimental Development and Innovations project `IT4Innovations National Supercomputing Center (LM2015070)'.

\bibliographystyle{elsarticle-num}
\biboptions{sort&compress}

\bibliography{bibliography}

\end{document}